\pdfoutput=1
\documentclass[twocolumn]{article}

\usepackage[a4paper,left=16mm,right=16mm,top=20mm,bottom=22mm]{geometry}

\usepackage[utf8]{inputenc}
\usepackage[T1]{fontenc}

\usepackage{amsmath,amsthm,amssymb}
\usepackage{newtxtext,newtxmath}

\usepackage{graphicx}
\usepackage{booktabs,tabularx}
\usepackage{array} 
\newcolumntype{Y}{>{\arraybackslash}X} 
\usepackage{float}
\usepackage{newfloat}
\DeclareFloatingEnvironment[name={Supplementary Figure}]{suppfigure}
\usepackage{tikz}
\usetikzlibrary{fit,positioning,arrows.meta}
\usepackage{pgfplots}
\pgfplotsset{compat=1.18}
\usepgfplotslibrary{fillbetween}

\usepackage{subcaption}

\usepackage{algorithm}
\usepackage{algpseudocode}

\usepackage{nicefrac}
\usepackage{microtype}
\usepackage{enumitem}
\usepackage{lineno}

\usepackage{titlesec}
\titleformat{\section}{\large\scshape\raggedright}{\thesection}{0.6em}{}
\titleformat{\subsection}{\normalsize\itshape\raggedright}{\thesubsection}{0.6em}{}
\titleformat{\subsubsection}{\normalsize\itshape\raggedright}{\thesubsubsection}{0.6em}{}
\titlespacing*{\section}{0pt}{1.5ex plus .5ex}{1.0ex plus .3ex}
\titlespacing*{\subsection}{0pt}{1.2ex plus .4ex}{0.7ex plus .2ex}
\titlespacing*{\subsubsection}{0pt}{1.0ex plus .3ex}{0.5ex plus .2ex}

\usepackage{needspace,etoolbox}
\preto\section{\needspace{4\baselineskip}}
\preto\subsection{\needspace{3\baselineskip}}
\preto\subsubsection{\needspace{2\baselineskip}}

\usepackage[numbers,square]{natbib}

\usepackage{xcolor}
\usepackage[colorlinks=true,allcolors=black]{hyperref}

\usepackage{sidecap}
\sidecaptionvpos{figure}{c}

\usepackage{xwatermark}
\newwatermark[firstpage,color=gray!90,angle=0,scale=0.28,xpos=0in,ypos=-5.15in]{*correspondence: \texttt{arthur.rondeau@unige.ch}}

\usepackage{authblk}

\title{Revealing Higher-Order Interactions in Complex Networks: A U.S. Diplomacy Case Study}
\author[1,2]{Arthur Rondeau\thanks{\texttt{arthur.rondeau@unige.ch}}}
\author[2]{Didier Wernli\thanks{\texttt{didier.wernli@unige.ch}}}
\author[1,2]{Roland Bouffanais\thanks{\texttt{roland.bouffanais@unige.ch}}}
\affil[1]{Department of Computer Science, University of Geneva}
\affil[2]{Global Studies Institute, University of Geneva}

\begin{document}

\twocolumn[ 
  \begin{@twocolumnfalse} 

\maketitle

\begin{abstract}
    Although diplomatic communication has long been examined in the social sciences, its network structure remains underexplored. Using the U.S. diplomatic cables released by WikiLeaks in 2010 as a case study, we adopt a network-science perspective. We represent diplomatic interactions as a hypergraph and develop a general, random-walk–based pipeline to evaluate this representation against traditional pairwise graphs. 
    We further evaluate the pipeline on legislative co-sponsorship and organizational-email data, finding improvements and empirical evidence that clarifies when hypergraph modeling is preferable to pairwise graphs.
    Overall, hypergraphs paired with appropriately specified random-walk dynamics more faithfully capture higher-order, group-based interactions, yielding a richer structural account of diplomacy and superior performance on interaction-prediction tasks that enables inferring new diplomatic relationships from existing patterns.
\end{abstract}
\vspace{0.35cm}

  \end{@twocolumnfalse} 
] 



\section{Introduction}
Over recent years, diplomatic communication has been extensively examined within international relations from a social science perspective \cite{IdeaOfTerror2020, StruggleForMinds2020, HumanRightsVsNationalInterests2019}.
One prominent case is the \textit{CableGate} leak, the release of US diplomatic cables by WikiLeaks in 2010.
While the \textit{CableGate} disclosures received widespread media coverage and continue to attract public attention \cite{Cablegate_Coverage}, they have been only sparsely analyzed in political science research \cite{michael_whos_2015} and, to date, never through the lens of network science.
In this work, we aim to fill this gap by introducing a network-based framework to analyze the US diplomatic relations. 
In particular, we propose a framework based on higher-order networks, especially hypergraphs (i.e. graphs where edges can connect more than two nodes) combined with random walk dynamics,
to better capture the complexity of diplomatic relationships.
Hypergraphs are particularly well suited to this domain, as diplomatic interactions often involve multiple actors simultaneously, 
forming relationships that cannot be adequately represented by traditional pairwise graph models. 
In recent years, higher-order networks have been the subject of extensive study. Particularly hypergraphs and simplicial complexes, 
offering powerful tools for analyzing systems characterized by group interactions beyond pairs 
\cite{battiston_networks_2020, battiston_physics_2021}. 
Hypergraphs naturally model diverse real-world systems, 
ranging from co-authorship networks where hyperedges denote research groups \cite{lung_hypergraph_2018}, 
to group conversations in social contexts \cite{patania_shape_2017}, 
and ecological interactions among multiple species \cite{golubski_ecological_2016}. 
Unlike simplicial complexes, which require that higher-order interactions imply all lower-order links, 
hypergraphs directly encode multiway interactions without imposing such constraints \cite{courtney_generalized_2016}.
In all these cases, it is assumed that interactions cannot be decomposed into independent pairwise links. 
However, little work has been performed to actually question the relevance of a higher-order framework over a classical graph when dealing with new data.
Thus, in addition of studying US diplomatic relationships, we perform a benchmark of our framework on widely used hypergraphs datasets to quantify the advantage of hypergraphs over classical graphs. 
To evaluate this framework, we design self-supervised tasks random walks based focused on detecting spurious interactions (i.e. hyperedges) and predicting new ones.

\paragraph{}  
Within this framework, we focus on the \textit{Edge Dependent Vertex Weight} (EDVW) hypergraph model \cite{chitra_random_2019}, 
where a vertex’s weight varies depending on the hyperedge it belongs to.
This design captures the fact that an individual’s importance or influence can differ across group contexts 
—for instance, depending on the topic of discussion or the composition of the group. 
EDVW hypergraphs thus allow for asymmetric and context-dependent influence within group interactions, 
going beyond what standard hypergraphs can represent. 
This added flexibility makes EDVW hypergraphs especially well suited for modeling the complexity of real-world group dynamics. 
Accordingly, we adopt the EDVW hypergraph framework to represent U.S. diplomatic relationships. 

\paragraph{}
In the other hand, considering hypergraph as a standalone entity, i.e. without considering the dynamics at state could be less informative at best or even misleading as stated in \cite{neuhauser_multi-body_2019} where 
linear processes on edge-independent vertex-weighted hypergraphs (EIVW) (i.e. each node contributes equivalently to all its hyperedges) can be equivalently represented on projected undirected weighted graphs. Therefore, to fully exploit the advantages of higher-order networks, it is essential to focus on the dynamics that unfold within these structures. 
Such dynamics include opinion formation \cite{chu_density_2023, hickok2022bounded, neuhauser_multi-body_2019}, contagion processes \cite{xu_dynamics_2022,arruda_social_2019, arruda_contagion_2024}, synchronization phenomena \cite{lucas_multiorder_2020}. 
Here, we concentrate on random walk dynamics as a means to probe network structure.
Random walks are especially appealing in the context of diplomatic communication, 
as they provide a natural proxy for the flow of information between embassies and consulates while preserving interpretability. 
Previous work has established the theoretical foundations of random walks on EDVW hypergraphs 
and their equivalence to projected-graph walks under certain conditions \cite{chitra_random_2019}.
Related studies have examined random walks on edge-independent vertex-weighted (EIVW) hypergraphs \cite{carletti_random_2020}, 
explored their role in community detection \cite{carletti_random_2021}, hyperedge prediction \cite{xu_hyperlink_2023}, 
and investigated the interplay between network architecture and random walk dynamics \cite{eriksson_how_2021}.
However, these studies are restricted to Markovian dynamics and, crucially, to EIVW hypergraphs; consequently, the random-walk dynamics admit an equivalent formulation on the projected pairwise graph \cite{chitra_random_2019}, failing to exploit higher-order interactions.
Accordingly, we employ the EDVW hypergraph framework not only for its inherent flexibility but also to allow the random-walk dynamics to capture the full potential of higher-order interactions. 
Particular attention is devoted to contrasting the roles of Markovian and non-Markovian processes in shaping these dynamics.

\paragraph{}
In this context, our contribution is thus threefolds:
(i) we propose a pipeline to evaluate the effectiveness of higher-order networks by comparing hypergraphs to traditional pairwise graphs across various datasets and tasks, highlighting contexts where hypergraphs may or may not be advantageous,
(ii) we propose a novel non-Markovian random walk, named \textit{Hyperwalk}, specifically designed for EDVW hypergraphs, allowing to fully leverage potential of higher-order interactions, and
(iii) we illustrate the framework’s utility by showing that U.S. diplomatic communication relationships exhibit higher-order interactions; separately, we demonstrate that appropriately specified random-walk dynamics can predict relationship formation.

\section{Methods}
\subsection{Datasets Collection \& Hypergraph Formulation}
\subsection{EDVW Hypergraph}
Similarly as introduced in \cite{chitra_random_2019}, we define an hypergraph with edge-dependent vertex weights (EDVW) as follows:
Let $\mathcal{H} = (V, E, \omega, \gamma)$ be a hypergraph with edge-dependent vertex weights.
We define $E(v) = \{ e \in E : v \in e \}$ to be the hyperedges
incident to a vertex $v$, and $E(u,v) = \{ e \in E : u \in e, v \in e \}$ 
to be the hyperedges incident to both vertices $u$ and $v$.
Let $d(v) = \sum_{e \in E(v)} \omega(e)$ denote the degree of vertex $v$, 
and let $\delta(e) = \sum_{v \in e} \gamma_e(v)$ denote the degree of 
hyperedge $e$. 

The \textit{vertex-weight matrix} $R$ of a hypergraph with edge-dependent 
vertex weights $H = (V,E,\omega,\gamma)$ is an $|E| \times |V|$ matrix 
with entries $R(e,v) = \gamma_e(v)$, and the \textit{hyperedge-weight 
matrix} $W$ is a $|V| \times |E|$ matrix with $W(v,e) = \omega(e)$ if 
$v \in e$, and $W(v,e) = 0$ otherwise. 

The vertex-degree matrix $D_V$ is a $|V| \times |V|$ diagonal matrix with 
entries $D_V(v,v) = d(v)$, and the hyperedge-degree matrix $D_E$ is a 
$|E| \times |E|$ diagonal matrix with entries $D_E(e,e) = \delta(e)$.

\subsubsection{US CableGate Dataset}
Our study is focused on diplomatic communication, especially on the US CableGate case study. 
The US CableGate is a collection of 251,287 US diplomatic cables and documents from more than 250 US embassies and consulates around the world. 
They can be divided into six categories: \textit{secrets not to be shared with non-Americans}, \textit{secrets}, \textit{confidential information not to be shared with non-Americans}, 
\textit{confidential information}, and \textit{unclassified information for official use and unclassified information} and finally \textit{unclassified}. No cable was classified as \textit{top secret} on the classification scale \cite{ExecOrder13526}.

In particular, we looked at relationships between US embassies and consulates between early 2000 and 2012, a communication will be named as a \textit{Cable} for the rest of the text. We processed 181000 cables, scrapped from WikiLeaks website.
Each communication contains the following set of information \{sender, receivers, timestamp, cable title, text content of cable, cable classification\}. 
Thus, our work does not rely on the content of the cables for legal and ethical reasons, but only on the metadata of \{sender, receivers, timestamp\}. 
Each cable is represented as a hyperedge, connecting multiple nodes (embassies and consulates). The weight of each hyperedge $w(e)$ reflects the importance of a communication (here based on the number of entities involded), while the edge-dependent vertex weights $\gamma_e(v)$ capture the varying significance of each embassy or consulate within different cables (here based on the role of the entity in the communication, as \textit{sender} or \textit{receiver}).
An overview of the EDVW hypergraph formulation is resumed ~Tables \ref{tab:cables-raw},\ref{tab:edvw-city},\ref{tab:edvw-country}. 

\begin{table}[htbp]
    \centering
    \caption{Examples of raw cables used to construct the EDVW hypergraph.}
    \label{tab:cables-raw}
    \begin{tabularx}{\linewidth}{@{}l l Y@{}}
    \toprule
    Cable & From (mission) & To (missions) \\
    \midrule
    C1 & Embassy Tunis  & Embassy Ankara; Consulate Istanbul; Embassy Bruxelles \\
    C2 & Embassy Rabat  & Embassy Tunis \\
    \bottomrule
    \end{tabularx}
    
    \vspace{0.5ex}
    \footnotesize\emph{Note.} Each cable induces one hyperedge \(e\).
    \end{table}
    
    \begin{table}[htbp]
    \centering
    \caption{City-level EDVW Hypergraph.}
    \label{tab:edvw-city}
    \begin{tabularx}{\linewidth}{@{}l Y c c@{}}
    \toprule
    Cable & \(y_e(v)\) by city & \(w_e\) & \(\displaystyle \sum_v y_e(v)\) \\
    \midrule
    C1 & \{Tunis: 2,\; Ankara: 1,\; Istanbul: 1,\; Bruxelles: 1\} & 4 & 5 \\
    C2 & \{Rabat: 1,\; Tunis: 2\} & 2 & 3 \\
    \bottomrule
    \end{tabularx}
    
    \vspace{0.5ex}
    \footnotesize\emph{Notes.} \(y_e(v)\) is the edge-dependent vertex weight with \(y_e(v)=2\) if \(v\) is the sender of cable \(e\), else \(y_e(v)=1\); 
    \(w_e=\lvert\{v: y_e(v)>0\}\rvert\) counts distinct cities; the last column reports the multiset size \(\sum_v y_e(v)\).
    \end{table}
    
    \begin{table}[htbp]
    \centering
    \caption{Country-level EDVW Hypergraph.}
    \label{tab:edvw-country}
    \begin{tabularx}{\linewidth}{@{}l Y c c@{}}
    \toprule
    Cable & \(y_e(v)\) by country & \(w_e\) & \(\displaystyle \sum_v y_e(v)\) \\
    \midrule
    C1 & \{Tunisia: 2,\; T\"urkiye: 1,\; Belgium: 1\} & 3 & 4 \\
    C2 & \{Morocco: 1,\; Tunisia: 2\} & 2 & 3 \\
    \bottomrule
    \end{tabularx}
    
    \vspace{0.5ex}
    \footnotesize\emph{Notes.} \(y_e(v)\) is the edge-dependent vertex weight with \(y_e(v)=2\) if \(v\) is the sender of cable \(e\), else \(y_e(v)=1\); 
    \(w_e\) counts distinct countries in \(e\); the last column reports the multiset size \(\sum_v y_e(v)\).
    \end{table}

\subsubsection{Common Hypergraphs Datasets}

The other datasets used are ones commonly used in the literature. \textit{Email-Eu} \cite{yin2017local}, \textit{Email-Enron} \cite{Benson-2018-simplicial}, \textit{Senate-Bills} \cite{Benson-2018-simplicial, Fowler-2006-connecting, Fowler-2006-cosponsorship}.
See Appendix~\ref{appendix:common-datasets-formation} for more details on the hypergraph formulation.
\newpage

\subsection{Random Walks}
In this section, we introduce all the random walks used in the following experiments and in particular \textit{hyperwalk} the non-Markovian random walk. See Appendix~\ref{appendix:random-walk-properties} for random walks properties among all datasets.
The important point here in that the random walks are defined on EDVW hypergraphs contexts.

\subsubsection{EDVW Hypergraph Markovian Random Walk}
Let $\mathcal{H} = (V, E, \omega, \gamma)$ be a hypergraph with edge-dependent vertex weights. We first define a random walk on $\mathcal{H}$ based on \cite{chitra_random_2019}. At time $t$, a random walker at vertex $v_t$ proceeds as follows:

\begin{enumerate}
    \item Pick an edge $e$ containing $v$, with probability $\frac{\omega(e)}{d(v)}$.
    \item Pick a vertex $w \in e$, with probability $\frac{\gamma_e(w)}{\delta(e)}$.
    \item Move to vertex $v_{t+1} = w$, at time $t+1$.
\end{enumerate}

The random walk on a hypergraph with edge-dependent vertex weights $\mathcal{H} = (V, E, \omega, \gamma)$ is a Markov chain on $V$ with transition probabilities
\begin{equation}
    p_{v,w} = \sum_{e \in E(v)} \left( \frac{\omega(e)}{d(v)} \right) \left( \frac{\gamma_e(w)}{\delta(e)} \right),
\end{equation}
where $E(v)$ denotes the set of hyperedges containing vertex $v$.

The probability transition matrix $P$ of the random walk on $\mathcal{H}$ is the $|V| \times |V|$ matrix with entries $P(v, w) = p_{v,w}$ and can be written in matrix form as:
\begin{equation}
    P = D_V^{-1} W D_E^{-1} R,
\end{equation}
$W$ is the vertex–hyperedge weight matrix with $W(v,e) = \omega(e)$ if $v \in e$, else $0$,
and $R$ is the hyperedge–vertex weight matrix with $R(e,v) = \gamma_e(v)$.
$d(v) = \sum_{e \in E(v)} \omega(e)$  denote the degree of vertex $v$, and let $\delta(e) = \sum_{v \in e} \omega_{e}(v)$ denote the degree of hyperedge $e$.
$D_V$ is the vertex degree matrix with $D_V(v,v) = d(v)$, $D_E$ is the hyperedge degree matrix with $D_E(e,e) = \delta(e)$.

\subsubsection{EDVW Hypergraph Non-Markovian Random Walk}
Let $\mathcal{H} = (V, E, \omega, \gamma)$ be a hypergraph with edge-dependent vertex weights. We introduce a non-markovian random walk on $\mathcal{H}$. At time $t$, a random walker at vertex $v_t$ proceeds as follows:

\begin{enumerate}
    \item Pick an edge $e$ containing $v$, with probability $\frac{\omega(e)}{d(v)}$.
    \item Pick a vertex $w \in e$, with probability $\frac{\gamma_e(w)}{\delta(e)}$.
    \item Move to vertex $v_{t+1} = w$, at time $t+1$.
    \item Restart from step $1.$ with a probability $1 - \frac{(|e|-2)}{|e|}$.
\end{enumerate}
The key difference with the Markovian random walk is the restart probability at each step, which depends on the size of the hyperedge $|e|$ chosen at step 1.
This restart probability is designed to capture the intuition that in larger groups (i.e., larger hyperedges), the likelihood of continuing to interact within the same group increases, in the same spirit as the time based community detection in hypergraphs \cite{carletti_random_2021}.
Thus, the next step depends on the information carried by the previous edge. 
The random walk is Markovian when the state is defined as $(v_t, e_t)$, 
where $e_t$ denotes the edge containing $v_t$ at time $t$, 
but becomes non-Markovian when the state space is restricted to $(v_t)$ alone.

\textit{Transition Probabilities Approximation}\\
Since the walk is non-Markovian on the states $(v_t)$, for the experiments relying on a \textit{vertex-to-vertex} transition probability matrix, we resorted to a Monte Carlo estimation of the transition probabilities. Specifically, we proceed as follows:
\begin{enumerate}
    \item For each vertex $v$, generate $N$ random walk paths of length $K$ according to the dynamics defined previously.
    \item Approximate the transition matrix at each step by computing the empirical frequencies of the transitions observed across the $N$ sampled paths.
\end{enumerate}

\subsubsection{Clique-Graph Markovian Random Walk}
Let $\mathcal{H} = (V, E, \omega, \gamma)$ be a hypergraph with edge-dependent vertex weights, where the vertex weights are normalized such that $\rho_e = 1$ for all hyperedges $e$ \cite{chitra_random_2019}. Let $G_{\mathcal{H}}$ denote the clique graph of $\mathcal{H}$, with edge weights defined by:

\begin{equation}
    w_{u,v} = \sum_{e \in E(u,v)} \frac{\omega(e) \, \gamma_e(u) \, \gamma_e(v)}{\delta(e)},
\end{equation}

where $E(u,v)$ is the set of hyperedges containing both $u$ and $v$, $\omega(e)$ is the weight of edge $e$, $\gamma_e(u)$ is the vertex weight of $u$ in $e$, and $\delta(e) = \sum_{v \in e} \gamma_e(v)$ is the degree of hyperedge $e$.
The transition matrix of the random walk on clique graph is then defined as $P \backslash P_{i,j} = \omega_{i,j}$.

\subsubsection{Hypergraph \& Graph Random Walks Equivalence}
The goal of this section is to establish the necessary conditions to prove that a random walk on an EDVW hypergraph (Markovian or approximation from non-Markovian) cannot be reduced to a random walk on a projected weighted undirected graph.
Knowing that every \textit{time-reversible Markov chains can be reduced to random walks on undirected graphs} \cite{lovasz1993random}, we needed to ensure the non-time-reversibility of the processes.
Thus, in order to prove the non-equivalence of the random walk on a hypergraph and the random walk on a projected/clique graph, we need to show that the random walk on the EDVW Hypergraph is not reversible, i.e. does not statisfy the detailed balance condition:
\begin{equation}
    \pi_{i} P_{i,j} = \pi_{j} P_{j,i} \quad \forall \, i,j \in V
\end{equation}
For more details, see Appendix~\ref{appendix:random-walk-properties}.

\subsection{Fake Hyperedge Detection Task}
The first task to evaluate the relevance of hypergraph framework is a fake detection task, that could be considered as an anomaly detection task.
The detection is a random walk based algorithm, inspired from the one performed on classical graphs \cite{liu_link_2010}.
The key point is a similarity matrix $S$ computed from the transition matrix $P$ of the random walk on the hypergraph, according to the following formula:
\begin{equation}
    S = \dfrac{1}{K}\sum_{k=1}^{K} P^{k}
\end{equation}
Each line $S_i$ of the similarity matrix $S$ could be interpreted as a feature vector of node $i$, capturing the structural context of node $i$ within $K$ steps of the random walk regarding all the other nodes.
In fact, each row $S_{i}$ is the distribution of probability to reach any other node $j$ from node $i$ within $K$ steps of the random walk by choosing randomly one of the $K$ walks.
From this similarity matrix, we can compute a score for each hyperedge $e$ in the hypergraph according to the following formula:
\begin{equation}
    S_e = 1 - \mathrm{JS}_{\pi_t}(S_{i_1}, \ldots, S_{i_t}) / \log_2 t,
    \end{equation}
        
    where $t = |e|$, $\pi_t = (1/t, \ldots, 1/t)$ and $e = \{ i_1, \ldots, i_t \}$.
\paragraph{}
This generalized Jensen–Shannon divergence (GJS) score based captures the non pairwise dependences of nodes within in hyperedge, as mentioned in \cite{xu_hyperlink_2023,eriksson_how_2021}. For more details on the scoring method, see Appendix~\ref{appendix:scoring-method}.
From the set of hyperedges, we sample a set of \textit{training} hyperedges $E^{T} $ and a set of \textit{probe} hyperedges noted $E^{P}$. 

\begin{figure}[h!]
    \centering
    \includegraphics[width=0.45\textwidth]{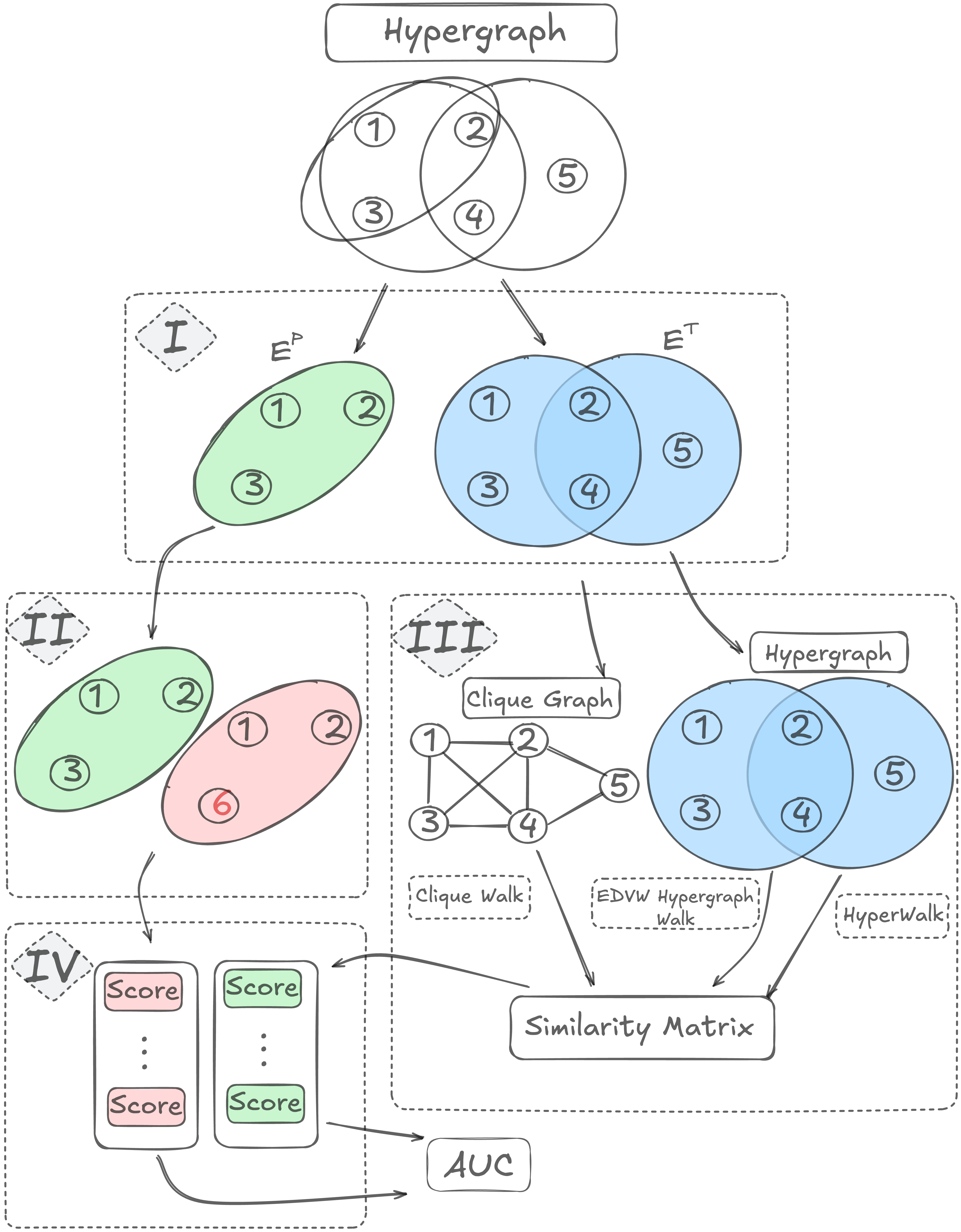} 
    \caption{Overview of the fake hyperedge detection procedure. In the step 2 of sampling fake hyperedges, the node $3$ has been replaced by node $6$ and hyperedge $\{1,2,6\}$ does not appear in $E$. The procedure is repeated $k$ times as part of a cross-validation scheme (number of folds $k=10$).}
    \label{fig:fake_detection_procedure}
\end{figure}

The following procedure is repeated $k$ times as part of a cross-validation scheme:

\begin{itemize}[leftmargin=1.5em, itemsep=2pt, topsep=2pt]
    \item \textbf{Step 1:} Construct a candidate edge set $E_c$ from the hypergraph $H(E,V,\gamma,\omega)$. This set is split into two disjoint subsets: $E^{T}$ (\textit{Training Set}) and $E^{P}$ (\textit{Probe Set}).  
    When sampling $E^{P}$, before extracting each hyperedge $e \in E^{P}$, we check if the remaining hypergraph from $E^{T}$ is still connected. In addition, all probe hyperedges in $E^{P}$ that are strict subsets of hyperedges in $E^{T}$ are discarded to ensure no data leak between the two sets.
    \item \textbf{Step 2:} The fake hyperedges set $E^{f}$ is generated from $E^{P}$ using different sampling strategies. For all strategies, we ensure that the hyperedges sampled do not already belong to the original set of hyperedges $E$.
    \begin{itemize}[leftmargin=2em, itemsep=1pt, topsep=1pt]
        \item \textbf{Alpha}: Replace a fraction $1-\alpha$ of the nodes in each probe hyperedge in $E^{P}$.
        \item \textbf{Degree-Matched}: Sample negative edges by matching node degrees to preserve structural consistency.
        \item \textbf{K-Replace}: Sample negative edges by randomly replacing the $K$ nodes of each hyperedge.
    \end{itemize}

    \item \textbf{Step 3:} On the remaining hypergraph defined by $E^{T}$, compute the similarity matrix $S$ with the different transition matrix of random walks described in \textit{3.2 Random Walks}.
    \item \textbf{Step 4:} Compute the GJS score of the fake and true hyperedges. Compute the AUC of the similarity scores of the hyperedges in $E^{P}$ and $E^{f}$. The AUC is computed using the following classic formula: $\dfrac{(n'' + 0.5n')}{n}$, where $n''$ is the number of true positive predictions, $n'$ is the number of ties in scores predictions, and $n$ is the total number of pair comparison between true and fake scores.
\end{itemize}
A summary of the procedure is illustrated in figure~\ref{fig:fake_detection_procedure}.

\subsection{Hyperedge Prediction Task}
Can we predict new embassy-consulate interactions from the observed ones? Can a hypergraph representation combined with an appropriate random-walk dynamics recover the latent structure of diplomatic relations?
To answer those questions, we perform a hyperedge prediction task.
The procedure is repeated $k$ times as part of a cross-validation scheme.
\begin{itemize}[leftmargin=1.5em, itemsep=2pt, topsep=2pt]
    \item \textbf{Step 1:} Construct a candidate edge set $E_c$ from the hypergraph $H(E,V,\gamma,\omega)$. This set is split into two disjoint subsets: $E^{T}$ (\textit{Training Set}) and $E^{P}$ (\textit{Probe Set}).  
    When sampling $E^{P}$, before extracting each hyperedge $e \in E^{P}$, we check if the remaining hypergraph from $E^{T}$ is still connected. In addition, all probe hyperedges in $E^{P}$ that are strict subsets of hyperedges in $E^{T}$ are discarded to ensure that no data leakage between the two sets.

    \item \textbf{Step 2:} The hyperedges to guess set $E^{g}$ is generated from $E^{P}$ using different sampling strategies over \texttt{n\_trials}.
        \begin{itemize}
        \item \textbf{Alpha}: Delete a fraction $1-\alpha$ of the nodes in each probe hyperedge in $E^{P}$.
        \end{itemize}
        \item \textbf{Step 3:} On the remaining hypergraph defined by $E^{T}$, compute the similarity matrix $S$ with the different transition matrix of random walks described in \textit{3.2 Random Walks}.
        \item \textbf{Step 4:} Guess the $K$ or $|e|(1-\alpha)$ missing nodes of each hyperedge in $E^{g}$ using the similarity matrix $S_{k}$, with $k \in \lbrack 1, K \rbrack$ or $k \in \lbrack 1, |e|(1-\alpha) \rbrack$. The guessed nodes are the ones maximizing the GJS similarity score with the remaining nodes of the hyperedge. Ensure Algorithm~\ref{alg:guess-refine} is defined.
\end{itemize}

\begin{figure}[h!]
    \centering
    \includegraphics[width=0.45\textwidth]{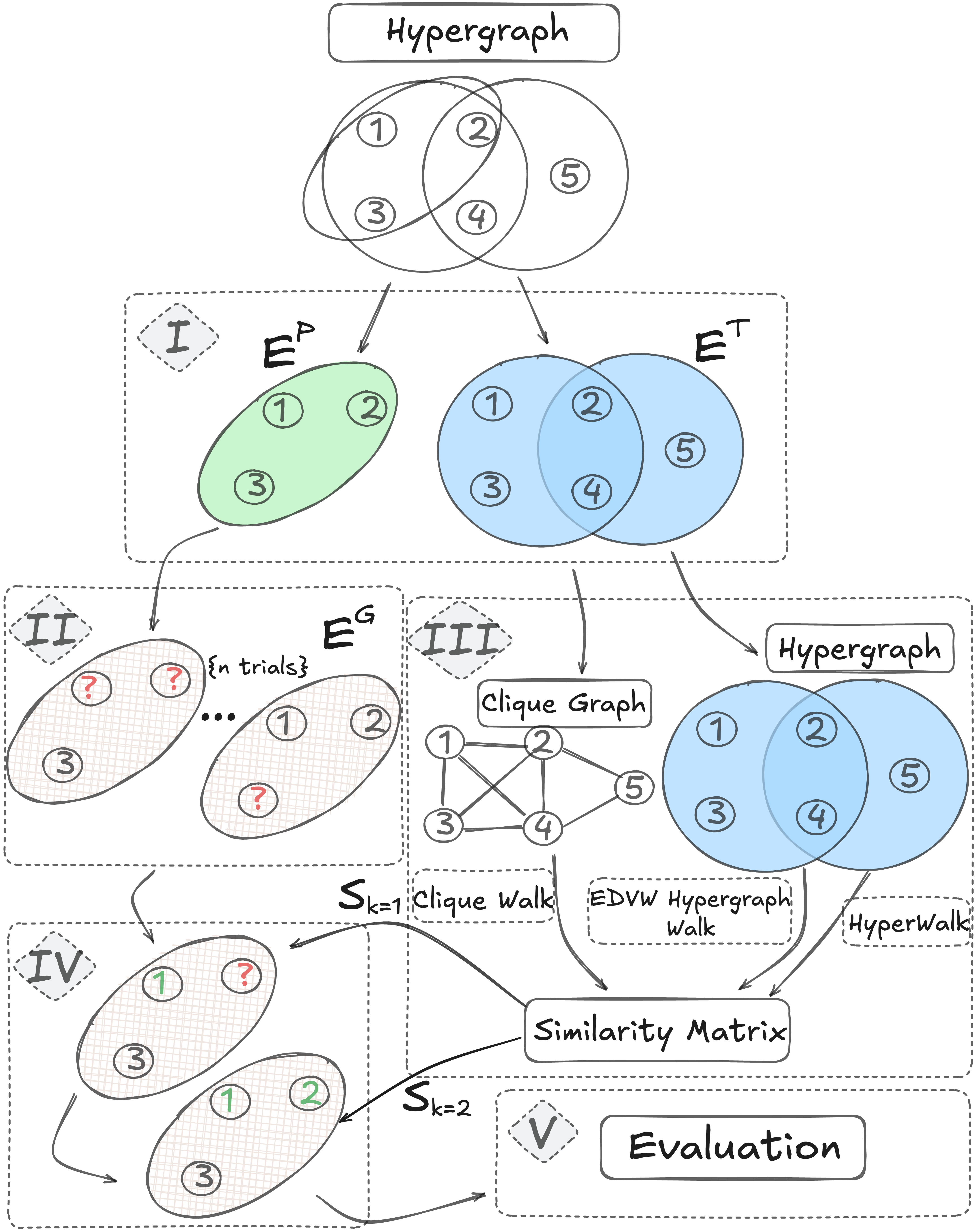} 
    \caption{Overview of the hyperedge prediction procedure. In the step 2 of sampling fake hyperedges, the node $3$ has been replaced by node $6$ and hyperedge $\{1,2,6\}$ does not appear in $E$. The procedure is repeated $k$ times as part of a cross-validation scheme (number of folds $k=10$).}
    \label{fig:hyperedge_prediction}
\end{figure}

A summary of the procedure is illustrated in figure~\ref{fig:hyperedge_prediction}.

\section{Results}
\subsection{Equivalence of Hypergraph and Graph Random Walks}
First, we verified the non-equivalence of the random walks on the EDVW hypergraph and on the projected/clique graph for all datasets.
From our results, all the random walks on the EDVW hypergraph are non-reversible (i.e. non-reducible to random walks on a projected graph). As expected, the random walk on EIVW hypergraph is reducible to a random walk on a projected graph.
We summarize the results in appendix Table~\ref{tab:markov-random-walk-properties}-\ref{tab:clique-random-walk-properties}.

\subsection{Fake Hyperedge Detection Task}
Tables~\ref{tab:best-AUC-US-Cables-City-TASK1}–\ref{tab:best-AUC-US-Cables-City-TASK1-degree2} report the best performance of the different \textit{Network Architecture + Random Walk} in \textit{TASK 1} (detection of fake hyperedges) in the US CableGate dataset at embassies and consulates level, using, respectively, the \textit{Alpha}, \textit{K-Replace} and \textit{Degree-Matching} sampling methods.
Examples of results with hypergraph on country scale are presented in Table~\ref{tab:best-AUC-US-Cables-Country-TASK1-alpha0.5}.
In Table~\ref{tab:best-AUC-senate-bills-TASK1-k2} are presented the performances of the different models and architectures on the \textit{Senate-Bills} dataset.
The rest of the results on the other datasets are presented in Appendix~\ref{appendix:common-datasets-results}.

\begin{table}[!ht]
  \centering
  \caption{Best AUC per method and size bin (fake sampling method \textit{Alpha-Value} with $\alpha=0.5$) on CableGate City dataset, reported as mean $\pm$ std of the per-fold maxima. AUC samples number $n=1000$.}
  \label{tab:best-AUC-US-Cables-City-TASK1} 
  \begin{tabular}{l l c}
    \toprule
    Hyperedge size & Method & Mean $\pm$ Std \\
    \midrule
    3–6 & EDVW\_clique & 0.8743 $\pm$ 0.0120 \\
    3–6 & EDVW\_hyper & 0.8678 $\pm$ 0.0146 \\
    3–6 & \textbf{hyperwalk} & \textbf{0.8915} $\pm$ \textbf{0.0172} \\
    \midrule
    \addlinespace[0.5ex]   
    7–10 & EDVW\_clique & 0.9656 $\pm$ 0.0073 \\
    7–10 & EDVW\_hyper & 0.9635 $\pm$ 0.0080 \\
    7–10 & \textbf{hyperwalk} & \textbf{0.9720} $\pm$ \textbf{0.0067} \\
    \midrule
    \addlinespace[0.5ex]
    11–71 & EDVW\_clique  & 0.9902 $\pm$ 0.0048 \\
    11–71 & EDVW\_hyper & 0.9900 $\pm$ 0.0032 \\
    11–71 & \textbf{hyperwalk}  & \textbf{0.9929} $\pm$ \textbf{0.0018} \\
    \bottomrule
  \end{tabular}
\end{table}
\begin{table}[!ht]
    \centering
    \caption{Best AUC per method and size bin (fake sampling method \textit{K-Replace} with $K=2$) on CableGate City dataset, reported as mean $\pm$ std of the per-fold maxima. AUC samples number $n=1000$.}
    \label{tab:best-AUC-US-Cables-City-TASK1-k2}
    \begin{tabular}{l l c}
      \toprule
      Hyperedge size & Method & Mean $\pm$ Std \\
      \midrule
      3–6 & EDVW\_clique & 0.8234 $\pm$ 0.0230 \\
      3–6 & EDVW\_hyper & 0.8185 $\pm$ 0.0218 \\
      3–6 & \textbf{hyperwalk} & \textbf{0.8386} $\pm$ \textbf{0.0199} \\
      \midrule
      \addlinespace[0.5ex]
      7–10 & EDVW\_clique & 0.8678 $\pm$ 0.0182 \\
      7–10 & EDVW\_hyper & 0.8701 $\pm$ 0.0163 \\
      7–10 & \textbf{hyperwalk} & \textbf{0.8825} $\pm$ \textbf{0.0094} \\
      \midrule
      \addlinespace[0.5ex]
      11–71 & EDVW\_clique & 0.8209 $\pm$ 0.0185 \\
      11–71 & EDVW\_hyper & 0.8164 $\pm$ 0.0153 \\
      11–71 & \textbf{hyperwalk} & \textbf{0.8276} $\pm$ \textbf{0.0144} \\
      \bottomrule
    \end{tabular}
  \end{table}
\begin{table}[!ht]
    \centering
    \caption{Best AUC per method and size bin(fake sampling method \textit{K Degree-Matched} with $K=2$) on CableGate City dataset, reported as mean $\pm$ std of the per-fold maxima. AUC samples number $n=1000$.}
    \label{tab:best-AUC-US-Cables-City-TASK1-degree2}
    \begin{tabular}{l l c}
      \toprule
      Hyperedge size & Method & Mean $\pm$ Std \\
      \midrule
      3–6 & EDVW\_clique & 0.5382 $\pm$ 0.0120 \\
      3–6 & EDVW\_hyper & 0.5462 $\pm$ 0.0107 \\
      3–6 & \textbf{hyperwalk} & \textbf{0.5517} $\pm$ \textbf{0.0095} \\
      \midrule
      \addlinespace[0.5ex]
      7–10 & EDVW\_clique & 0.5432 $\pm$ 0.0078 \\
      7–10 & EDVW\_hyper & 0.5491 $\pm$ 0.0108 \\
      7–10 & \textbf{hyperwalk} & \textbf{0.5522} $\pm$ \textbf{0.0060} \\
      \midrule
      \addlinespace[0.5ex]
      11–71 & EDVW\_clique & 0.5412 $\pm$ 0.0085 \\
      11–71 & EDVW\_hyper & 0.5387 $\pm$ 0.0058 \\
      11–71 & \textbf{hyperwalk} & \textbf{0.5549} $\pm$ \textbf{0.0084} \\
      \bottomrule
    \end{tabular}
  \end{table}


\begin{table}[!ht]
    \centering
    \caption{Best AUC per method and size bin (fake sampling method \textit{K-Replace} with $K=2$) on senate-bills dataset, reported as mean $\pm$ std of the per-fold maxima. The AUC is the mean over folds of the per-fold max AUC. For all size bins and folds, the number of AUC samples is $n=1000$.}
    \label{tab:best-AUC-senate-bills-TASK1-k2}
    \begin{tabular}{l l c}
      \toprule
      Hyperedge & Method & Mean $\pm$ Std \\
      \midrule
      3–6 & EDVW\_clique & 0.8380 $\pm$ 0.0214 \\
      3–6 & EDVW\_hyper & 0.8300 $\pm$ 0.0254 \\
      3–6 & \textbf{hyperwalk} & \textbf{0.8791} $\pm$ \textbf{0.0258} \\
      \midrule
      \addlinespace[0.5ex]
      7–10 & EDVW\_clique & 0.8126 $\pm$ 0.0270 \\
      7–10 & EDVW\_hyper & 0.8153 $\pm$ 0.0298 \\
      7–10 & \textbf{hyperwalk} & \textbf{0.8439} $\pm$ \textbf{0.0291} \\
      \midrule
      \addlinespace[0.5ex]
      11–99 & EDVW\_clique & 0.6625 $\pm$ 0.0053 \\
      11–99 & EDVW\_hyper & 0.7338 $\pm$ 0.0091 \\
      11–99 & \textbf{hyperwalk} & \textbf{0.7587} $\pm$ \textbf{0.0089} \\
      \bottomrule
    \end{tabular}
  \end{table}

On the US-Cables City dataset, the EDVW hypergraph paired with the non-Markovian \textit{hyperwalk} does not consistently outperform other \textit{Network Architecture + Random Walk} combinations across size bins and negative-sampling schemes.
We do, however, observe the expected improvement with hyperedge size: under \textit{Alpha} sampling, larger hyperedges contain more fake-node substitutions, making them easier to detect.

When the number of substituted nodes is fixed at $k=2$ (see Table~\ref{tab:best-AUC-US-Cables-City-TASK1-k2} and Table~\ref{tab:best-AUC-US-Cables-City-TASK1-degree2}), performance drops markedly for large hyperedges because the signal from fake nodes does not scale with size, making them harder to differentiate from true hyperedges. Under \textit{Degree-Matched Sampling} (Table~\ref{tab:best-AUC-US-Cables-City-TASK1-degree2}), performance flattens and becomes largely insensitive to hyperedge size, suggesting that the random walk cannot exploit structural inconsistencies between fake and true hyperedges.

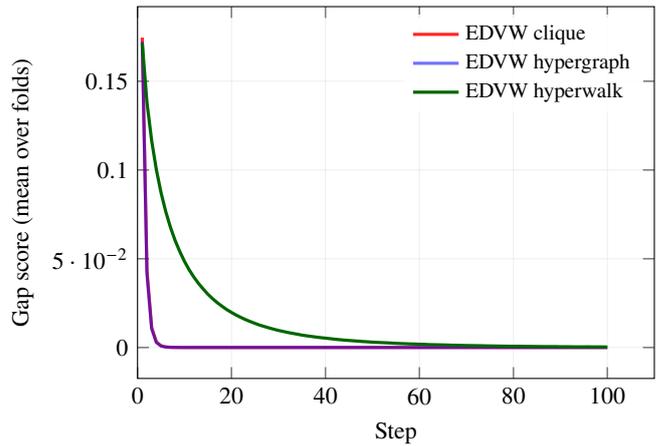
\begin{figure}[t]
    \centering
    \pgfplotsset{
      every axis plot/.append style={mark=none},
      tick style={semithick},
      label style={font=\small},
      tick label style={font=\small},
    }
    \begin{tikzpicture}
      \begin{axis}[
        width=0.8\columnwidth,
        height=0.58\columnwidth,
        scale only axis=true,
        xlabel={Step},
        ylabel={Gap score (mean over folds)},
        xmin=0,
        grid=both, grid style={opacity=0.25},
        legend style={font=\footnotesize, draw=none, fill=white, fill opacity=0.85, text opacity=1, at={(0.98,0.98)}, anchor=north east},
        legend cell align=left,
      ]
        \addplot[red, very thick, opacity=0.85] coordinates {(1.0,0.17455222835980866) (2.0,0.04331034125228495) (3.0,0.011263276883211782) (4.0,0.00311511172968205) (5.0,0.0009207611371185233) (6.0,0.0002871029877374733) (7.0,9.313997008646713e-05) (8.0,3.107712263778164e-05) (9.0,1.0563169362255377e-05) (10.0,3.6315213073339306e-06) (11.0,1.256951173982866e-06) (12.0,4.3683455448278864e-07) (13.0,1.522042580740247e-07) (14.0,5.3121031616449505e-08) (15.0,1.856108678105766e-08) (16.0,6.4906611817531405e-09) (17.0,2.271041725363343e-09) (18.0,7.949611779788454e-10) (19.0,2.783596995428962e-10) (20.0,9.749347465989749e-11) (21.0,3.415326301174256e-11) (22.0,1.1966351653822837e-11) (23.0,4.193281222761325e-12) (24.0,1.46960967659984e-12) (25.0,5.151124433979882e-13) (26.0,1.805707550724014e-13) (27.0,6.330416332498719e-14) (28.0,2.2196883178991204e-14) (29.0,7.782589837475442e-15) (30.0,2.7300085075138802e-15) (31.0,9.587655870985365e-16) (32.0,3.4011240939555104e-16) (33.0,1.237272389238803e-16) (34.0,4.3269687675937545e-17) (35.0,1.3490601054470782e-17) (36.0,2.95160511344092e-18) (37.0,9.875618085931152e-19) (38.0,-4.775753492831349e-19) (39.0,2.1446615504103222e-19) (40.0,2.0200319902174981e-19) (41.0,-1.1912217976203382e-19) (42.0,-2.689155452296787e-19) (43.0,-2.1454021215763904e-19) (44.0,4.943453078309247e-19) (45.0,-7.095893032330938e-20) (46.0,5.0997948377836906e-20) (47.0,-1.5016751747457195e-19) (48.0,-4.325329684604613e-20) (49.0,2.9063453959033155e-19) (50.0,-1.89783031786408e-19) (51.0,-1.2461310960350952e-19) (52.0,-9.971796407672577e-20) (53.0,2.221816378565838e-19) (54.0,7.712731171718531e-20) (55.0,8.030720106768731e-20) (56.0,-3.4512295387612176e-19) (57.0,1.5692485428862283e-19) (58.0,-1.3734879519081056e-19) (59.0,-1.537823281548967e-19) (60.0,-9.562299706491299e-20) (61.0,-4.570227362615842e-19) (62.0,-1.6699619887912386e-19) (63.0,3.4339853872265843e-19) (64.0,-3.273096622623279e-19) (65.0,2.78610267071435e-19) (66.0,-2.0139649942043413e-19) (67.0,1.5032312778035474e-19) (68.0,1.8114039641278415e-19) (69.0,2.478362591442926e-20) (70.0,-2.2005105278591987e-21) (71.0,-1.3975599967552029e-19) (72.0,3.054517296396306e-20) (73.0,-2.3098601807400583e-20) (74.0,8.379760128777072e-20) (75.0,-6.487907636494198e-19) (76.0,-7.453223618250256e-20) (77.0,1.3897478103334032e-19) (78.0,1.8717089832799496e-19) (79.0,5.266064859815862e-19) (80.0,-1.8743620511052554e-19) (81.0,-7.839325485194197e-20) (82.0,-2.907059920275935e-19) (83.0,4.634094360303137e-19) (84.0,7.064192505291608e-20) (85.0,4.783544084732603e-19) (86.0,3.3414799012807608e-19) (87.0,2.6371202950575636e-19) (88.0,9.843141946417058e-21) (89.0,-2.272926201758872e-19) (90.0,-2.0660480142226838e-19) (91.0,2.4336343086116847e-19) (92.0,-1.3957479847591242e-20) (93.0,2.0958343637986954e-19) (94.0,-3.9698224808068075e-19) (95.0,-3.0818577705148128e-19) (96.0,1.0569851309946813e-19) (97.0,-2.3525032282918503e-19) (98.0,-1.5457160706815334e-20) (99.0,-5.072607727798092e-20) (100.0,3.7421332092957706e-19)};
        \addlegendentry{EDVW clique}
        \addplot[blue, very thick, opacity=0.55] coordinates {(1.0,0.1721159052594687) (2.0,0.04115033322624244) (3.0,0.01041417591985697) (4.0,0.0028052868235886068) (5.0,0.0008065365879102501) (6.0,0.0002442683637702867) (7.0,7.69355697004644e-05) (8.0,2.4928334372685816e-05) (9.0,8.231322733166358e-06) (10.0,2.749828703193109e-06) (11.0,9.250112767286783e-07) (12.0,3.1245899000464086e-07) (13.0,1.0581976639837672e-07) (14.0,3.5898613044260365e-08) (15.0,1.2192390076122984e-08) (16.0,4.144291429983956e-09) (17.0,1.4094959524394815e-09) (18.0,4.795832008310031e-10) (19.0,1.6323241349678884e-10) (20.0,5.5572642304008516e-11) (21.0,1.892375534309852e-11) (22.0,6.44512454130053e-12) (23.0,2.1954484756555873e-12) (24.0,7.479583473769533e-13) (25.0,2.548517066650641e-13) (26.0,8.684704473384425e-14) (27.0,2.959901646744117e-14) (28.0,1.0088723634954284e-14) (29.0,3.43804327758072e-15) (30.0,1.1735698668414787e-15) (31.0,4.0333724029107574e-16) (32.0,1.4092410341991346e-16) (33.0,4.8783481233364566e-17) (34.0,1.5972803254687203e-17) (35.0,3.999596917392294e-18) (36.0,1.0743361146726517e-18) (37.0,-3.201056911471655e-19) (38.0,-1.8057666355172005e-19) (39.0,-4.0294934738016424e-19) (40.0,5.819681706278017e-19) (41.0,2.4986833023009675e-19) (42.0,-2.3916456934456244e-19) (43.0,-4.524988992056501e-20) (44.0,-1.7601718971124369e-19) (45.0,2.721930843572016e-19) (46.0,-1.4221691200328433e-19) (47.0,-3.100635266806145e-20) (48.0,1.876563531300453e-19) (49.0,-1.457335423423375e-19) (50.0,-5.3035787869996777e-20) (51.0,-7.119618675622958e-20) (52.0,-5.831233813143836e-20) (53.0,-7.551244356386376e-20) (54.0,4.2416862985917316e-20) (55.0,-4.7944077064203955e-20) (56.0,-1.5004566907050412e-19) (57.0,3.172812819468547e-19) (58.0,2.938699522824665e-19) (59.0,2.2355662865249995e-19) (60.0,1.2708581993632843e-19) (61.0,1.890841067592338e-19) (62.0,3.39062084169106e-19) (63.0,-1.7818365976915379e-19) (64.0,1.7597955213883847e-19) (65.0,-4.240377893751148e-19) (66.0,2.3702733496722174e-19) (67.0,-1.195964521317341e-19) (68.0,1.94415183365651e-19) (69.0,2.801513554866201e-20) (70.0,4.925884488445673e-19) (71.0,-3.19198433366907e-19) (72.0,1.7442365215560592e-19) (73.0,-1.056922370828755e-19) (74.0,-5.02211230311422e-20) (75.0,2.0643655084884558e-19) (76.0,1.182881217656619e-19) (77.0,1.2893413352943036e-19) (78.0,-4.8412916149508733e-20) (79.0,-4.019635654396166e-19) (80.0,-2.873565262136702e-19) (81.0,-1.9213517778256808e-19) (82.0,-1.0287188261458872e-19) (83.0,-4.0538034374862873e-19) (84.0,-1.2382656152431796e-20) (85.0,4.799849193561426e-19) (86.0,3.870713507988592e-20) (87.0,2.2687855507672667e-19) (88.0,-1.1046912380332326e-18) (89.0,2.4372984929605734e-19) (90.0,-9.003417650811778e-20) (91.0,-1.2202837363862904e-19) (92.0,-2.90457014199564e-20) (93.0,7.964338695159756e-21) (94.0,6.880566347266052e-20) (95.0,-1.6329221795082432e-19) (96.0,-1.7450133730799702e-19) (97.0,-3.195211675431786e-19) (98.0,-3.636326263764374e-20) (99.0,1.7105579685063673e-19) (100.0,9.079059904558464e-20)};
        \addlegendentry{EDVW hypergraph}
        \addplot[green!40!black, very thick] coordinates {(1.0,0.17136303575738765) (2.0,0.13808266885640041) (3.0,0.11651082785556556) (4.0,0.10017372862808975) (5.0,0.08716899177073358) (6.0,0.07657092626487778) (7.0,0.06771973831584165) (8.0,0.06026382645000801) (9.0,0.05399506375634439) (10.0,0.04845845831605018) (11.0,0.043695579497294336) (12.0,0.03961979044315007) (13.0,0.03601140211891788) (14.0,0.032879112663671016) (15.0,0.02992411607811308) (16.0,0.027424106311175363) (17.0,0.025182407112202564) (18.0,0.023198383448247827) (19.0,0.021305147501046458) (20.0,0.019732058904235905) (21.0,0.018225429414465744) (22.0,0.01692621659917868) (23.0,0.015671452612055364) (24.0,0.014596740148204324) (25.0,0.01356417600735518) (26.0,0.01258636548694223) (27.0,0.011801590809686563) (28.0,0.011006208750617103) (29.0,0.01028776865001146) (30.0,0.009593597167696399) (31.0,0.009024336813510165) (32.0,0.008444263188475033) (33.0,0.007953758209038146) (34.0,0.007469632223843436) (35.0,0.0069700469910927865) (36.0,0.00657622982119104) (37.0,0.006221193911073263) (38.0,0.005845802732905623) (39.0,0.005530819089110124) (40.0,0.00525276791122296) (41.0,0.00495928944552002) (42.0,0.004641405758196787) (43.0,0.004391867450081977) (44.0,0.004142951494192157) (45.0,0.003938027280075634) (46.0,0.0037547141609020297) (47.0,0.0035572103860526233) (48.0,0.00334795200883856) (49.0,0.003189208818896923) (50.0,0.003003181150964713) (51.0,0.0028611720586018334) (52.0,0.0027461381472523635) (53.0,0.0026240777883080033) (54.0,0.00249358076199649) (55.0,0.0023322900366624165) (56.0,0.0022027982879031568) (57.0,0.0021106761800776235) (58.0,0.0019871392310836468) (59.0,0.0019397978602356332) (60.0,0.0018197106594102601) (61.0,0.0017779894005257367) (62.0,0.00161218707591207) (63.0,0.0015567939089119868) (64.0,0.00150965271676603) (65.0,0.0014301958753034833) (66.0,0.0013697211011294667) (67.0,0.00125941114940692) (68.0,0.0012422564015199433) (69.0,0.0011790711026684433) (70.0,0.00109789374736808) (71.0,0.0010682142883934101) (72.0,0.00101923978199023) (73.0,0.0009817118564140767) (74.0,0.0009118660190299534) (75.0,0.0008636107697105533) (76.0,0.0008287149147213333) (77.0,0.0007808963077239534) (78.0,0.0007348138892903667) (79.0,0.0007049803795798734) (80.0,0.0006798745387270133) (81.0,0.00068519842003876) (82.0,0.0006478354435920832) (83.0,0.0006054110240016133) (84.0,0.0005828323160640834) (85.0,0.0005486025458712333) (86.0,0.00051545925476465) (87.0,0.0004894220184483734) (88.0,0.0004887729466554734) (89.0,0.00048536207532921333) (90.0,0.0004227492297690833) (91.0,0.0004228621754459433) (92.0,0.00040841097308385) (93.0,0.00035476941994899995) (94.0,0.00034309476824263667) (95.0,0.0003572070038074233) (96.0,0.00038003978063649667) (97.0,0.0003182238307287) (98.0,0.00033387741953965335) (99.0,0.00031877653043689005) (100.0,0.00024856102271850737)};
        \addlegendentry{EDVW hyperwalk}
      \end{axis}
    \end{tikzpicture}
    \caption{Gap score (mean over folds) across steps for EDVW clique, EDVW hypergraph, and EDVW hyperwalk. Negative sampling method \textit{Alpha-Value} with $\alpha=0.5$. Dataset: US Cables City.}
    \label{fig:gap_over_steps_US_Cables_City}
  \end{figure}

On the senate-bills dataset (Table~\ref{tab:best-AUC-senate-bills-TASK1-k2}), the EDVW hypergraph with \textit{hyperwalk} outperforms other \textit{Network Architecture + Random Walk} combinations over the widest range of hyperedge sizes ($11-71$), indicating higher-order interactions that are better captured by a non-Markovian, hypergraph-based random-walk framework. Taken together, these findings highlight that while a hypergraph formulation may be conceptually natural for group interactions, its empirical advantage is strongly dataset-dependent.
In some settings, a pairwise graph formulation can perform just as well or even better (see email-eu results in Appendix~\ref{appendix:email-eu-results}), even when considering large hyperedges. Results on other methods/datasets are presented in Appendix~\ref{appendix:common-datasets-results}.
\paragraph{}
In figure ~\ref{fig:gap_over_steps_US_Cables_City} the evolution of the similarity score gap between true and fake hyperedges is presented in the $k$ steps of the random walk for the different combinations of \textit{ network architecture + random walk} on the US CableGate City dataset.
First, we observe that the \textit{Hyperwalk} is the only random walk that remains consistent with $k>1$, while all other combos decrease, suggesting that the non-Markovian random walk is better at capturing the higher order structure of the hypergraph.
The gap being similar accross the methods for the first step, enables to explain why the performances of the fake hyperedge detection task are similar accross the different combos of \textit{Network Architecture + Random Walk}, showing that the task could be solve by only scanning the first neighbors of the nodes in the hyperedge.
Based on those observations, we can actually assume that the \textit{Fake Detection Task} is not challenging enough to highlight the advantage of using a hypergraph framework over a classical graph, thus we implemented another task, harder than the fake hyperedge detection task, the hyperedge prediction task.

\subsubsection{Intruder Detection Task}
  
Before presenting the results of the \textit{Hyperedge Prediction Task}, we introduce an intermediate task, named \textit{Intruder Detection Task}.
In this section, we demonstrate how in addition to differentiating fake to true interactions, we can actually identify the number of intruders (i.e. fakes nodes) in a fake hyperedges.
To do so, based on the score gap between true and fake hyperedges, we performed a regression task to predict the number of intruders in a fake hyperedge.

\begin{figure}[ht]
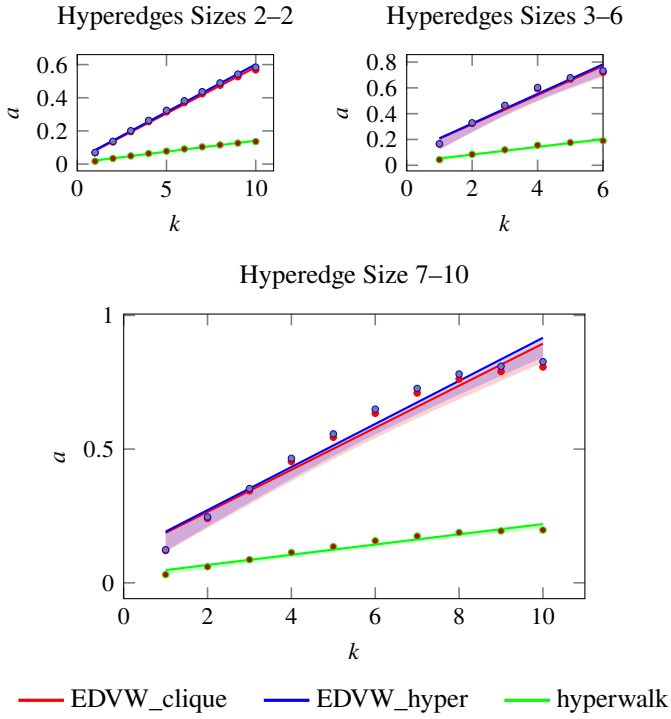

  \centering
  
  \pgfplotsset{
    every axis plot/.append style={mark=none},
    tick style={semithick},
    label style={font=\small},
    tick label style={font=\small},
  }
  
  \begin{minipage}{0.49\columnwidth}
  \centering

  \end{minipage}
  
  \caption{Linear regression coefficients ($a$) vs.\ number of intruders ($k$) across hyperedges sizes and models on US Cables City dataset.}
  \label{fig:regression_plots_US_Cables_City}
  \end{figure}

The model used to perform the regression is a simple exponential model, see Appendix~\ref{appendix:regression-method} for details.
Once the regressions performed for each combo of \textit{Network Architecture + Random Walk} on each range of size of hyperedges, for different values $n$ of intruders, we check whether from the estimated coefficients we could predict the number of intruders in a fake hyperedge.
On ~Figure \ref{fig:regression_plots_US_Cables_City} is presented the results of the regression task for the different combos of \textit{Network Architecture + Random Walk} on the US CableGate dataset.


We observe a clear linear dependency between the number of intruders and the parameter $a$ of the exponential model of the gap score (the choice of parameter $a$ over $b$ and $c$ is motivated by the fact that $a$ is the only parameter which was significantly different across the different number of intruders $n$ with a confidence of $95\%$).
Thus, from those results, we can actually conclude that not only differentiate fake to true interactions but in addition tell the number of intruders, by using either a hypergraph formulation or a classical graph.

\subsection{Hyperedge Prediction Task}
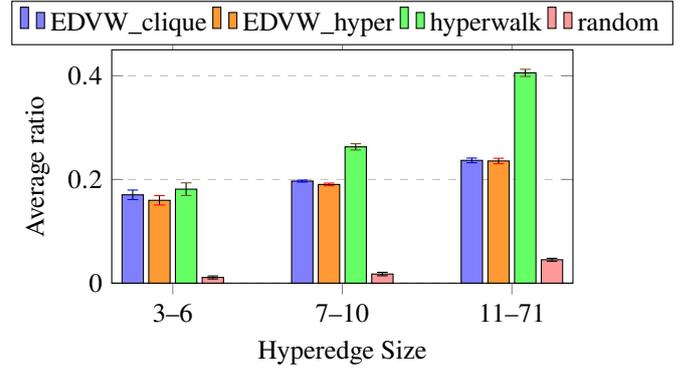
\begin{figure}[h!]
    \centering
    \begin{tikzpicture}
    \begin{axis}[
        ybar,
        width=0.9\linewidth,
        height=0.55\linewidth,
        bar width=8pt,
        enlarge x limits=0.18,
        ymin=0,
        ymax=0.45, 
        ylabel={Average ratio},
        xlabel={Hyperedge Size},
        xtick=data,
        symbolic x coords={3--6,7--10,11--71},
        legend style={at={(0.5,1.02)},anchor=south,legend columns=-1},
        ymajorgrids=true,
        grid style={dashed},
        error bars/y dir=both,
        error bars/y explicit,
    ]

    \addplot+[draw=black, fill=blue!50] coordinates {
      (3--6, 0.1704) +- (0, 0.0092)
      (7--10, 0.1969) +- (0, 0.0019)
      (11--71, 0.2368) +- (0, 0.0045)
    };
    \addlegendentry{EDVW\_clique}

    \addplot+[draw=black, fill=orange!80] coordinates {
      (3--6, 0.1598) +- (0, 0.0090)
      (7--10, 0.1904) +- (0, 0.0026)
      (11--71, 0.2356) +- (0, 0.0053)
    };
    \addlegendentry{EDVW\_hyper}

    \addplot+[draw=black, fill=green!60] coordinates {
      (3--6, 0.1813) +- (0, 0.0122)
      (7--10, 0.2629) +- (0, 0.0062)
      (11--71, 0.4056) +- (0, 0.0069)
    };
    \addlegendentry{hyperwalk}

    \addplot+[draw=black, fill=red!40] coordinates {
      (3--6, 0.0110) +- (0, 0.0029)
      (7--10, 0.0176) +- (0, 0.0033)
      (11--71, 0.0451) +- (0, 0.0028)
    };
    \addlegendentry{random}

    \end{axis}
    \end{tikzpicture}
    \caption{US Cables City dataset performances by size bin for overall hyperedges (mean $\pm$ std across folds), $\alpha=0.5$.}
    \label{fig:TASK2-US-cables-city-overall-ratio}\end{figure}

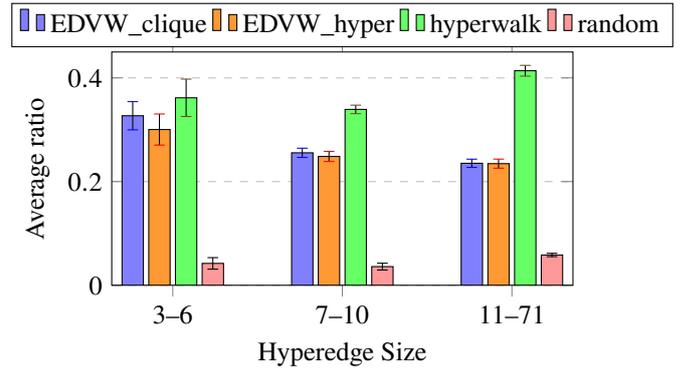
\begin{figure}[h!]
    \centering
    \begin{tikzpicture}
    \begin{axis}[
        ybar,
        width=0.9\linewidth,
        height=0.55\linewidth,
        bar width=8pt,
        enlarge x limits=0.18,
        ymin=0,
        ymax=0.45, 
        ylabel={Average ratio},
        xlabel={Hyperedge Size},
        xtick=data,
        symbolic x coords={3--6,7--10,11--71},
        legend style={at={(0.5,1.02)},anchor=south,legend columns=-1},
        ymajorgrids=true,
        grid style={dashed},
        error bars/y dir=both,
        error bars/y explicit,
    ]

    \addplot+[draw=black, fill=blue!50] coordinates {
      (3--6, 0.3270) +- (0, 0.0272)
      (7--10, 0.2555) +- (0, 0.0089)
      (11--71, 0.2353) +- (0, 0.0080)
    };
    \addlegendentry{EDVW\_clique}

    \addplot+[draw=black, fill=orange!80] coordinates {
      (3--6, 0.3004) +- (0, 0.0300)
      (7--10, 0.2485) +- (0, 0.0097)
      (11--71, 0.2346) +- (0, 0.0088)
    };
    \addlegendentry{EDVW\_hyper}

    \addplot+[draw=black, fill=green!60] coordinates {
      (3--6, 0.3616) +- (0, 0.0361)
      (7--10, 0.3392) +- (0, 0.0082)
      (11--71, 0.4138) +- (0, 0.0104)
    };
    \addlegendentry{hyperwalk}

    \addplot+[draw=black, fill=red!40] coordinates {
      (3--6, 0.0422) +- (0, 0.0111)
      (7--10, 0.0360) +- (0, 0.0066)
      (11--71, 0.0584) +- (0, 0.0033)
    };
    \addlegendentry{random}

    \end{axis}
    \end{tikzpicture}
    \caption{US Cables City dataset performances by size bin for novel hyperedges (mean $\pm$ std across folds), $\alpha=0.5$.}
    \label{fig:TASK2-US-cables-city-novel-ratio}\end{figure}

\begin{figure}[h!]
    \centering
    \begin{tikzpicture}
    \begin{axis}[
        ybar,
        width=\linewidth,
        height=0.55\linewidth,
        bar width=8pt,
        enlarge x limits=0.18,
        ymin=0,
        ymax=0.40,
        ylabel={Average ratio},
        xlabel={Size bin},
        xtick=data,
        symbolic x coords={2--2,3--6,7--10,11--99},
        legend style={at={(0.5,1.12)},anchor=south,legend columns=-1},
        ymajorgrids=true,
        grid style={dashed},
        error bars/y dir=both,
        error bars/y explicit,
    ]
    
    \addplot+[draw=black, fill=blue!50] coordinates {
        (3--6, 0.0955) +- (0, 0.0707)
        (7--10, 0.0875) +- (0, 0.0215)
        (11--99, 0.1366) +- (0, 0.0228)
    };
    \addlegendentry{EDVW\_clique}
    
    \addplot+[draw=black, fill=orange!80] coordinates {
        (3--6, 0.1197) +- (0, 0.0632)
        (7--10, 0.0890) +- (0, 0.0244)
        (11--99, 0.2289) +- (0, 0.0240)
    };
    \addlegendentry{EDVW\_hyper}
    
    \addplot+[draw=black, fill=green!60] coordinates {
        (3--6, 0.0825) +- (0, 0.0757)
        (7--10, 0.1241) +- (0, 0.0290)
        (11--99, 0.3127) +- (0, 0.0093)
    };
    \addlegendentry{hyperwalk}
    
    \addplot+[draw=black, fill=red!40] coordinates {
        (3--6, 0.0558) +- (0, 0.0435)
        (7--10, 0.0403) +- (0, 0.0110)
        (11--99, 0.0840) +- (0, 0.0021)
    };
    \addlegendentry{random}
    
    \end{axis}
    \end{tikzpicture}
    \caption{Senate-Bills dataset performances by size bin for novel hyperedges (mean $\pm$ std across folds), $\alpha=0.5$.}
    \label{fig:TASK2-senate-bills-novel-ratio}
  \end{figure}
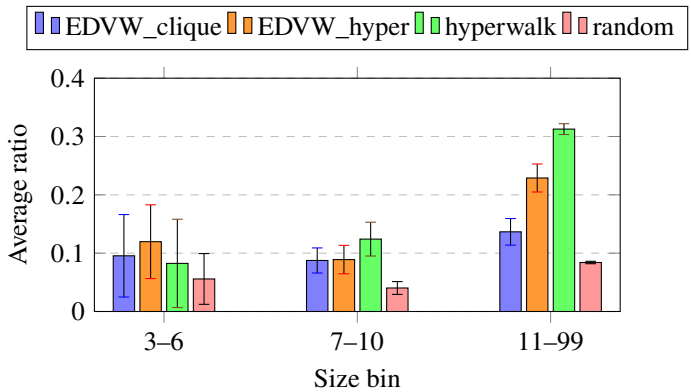

Here are presented the performances of the different combos \textit{Network Architecture + Random Walk} on the \textit{TASK 2} of hyperedge prediction for the US CableGate dataset at embassies and consulates level (city level) and for the senate-bills dataset.
In figure ~\ref{fig:TASK2-US-cables-city-overall-ratio} the overhaul ratio of corrects guessed nodes over the theoretical amount, in ~\ref{fig:TASK2-US-cables-city-novel-ratio},~\ref{fig:TASK2-senate-bills-novel-ratio}, the ratio of the \textit{novel} interactions (i.e set \textit{preserved + correctly guessed} that is not a subset of any hyperedge in $E^{T}$)
and finally ~\ref{fig:TASK2-US-cables-city-seen-ratio} of the seen interactions (i.e. it exists at least one identical set \textit{preserved + correctly guessed} that is a subset of a hyperedge in $E^{T}$). See Appendix~\ref{appendix:hyperedge-prediction-results-US-cables-city} for more details on the results presentation.
The motivation to divide the results into different categories is to better understand how the different combos of \textit{Network Architecture + Random Walk} can capture the underlying structure of diplomatic relationships, while ensuring that the random walk on the $E^{T}$  actually captured the structure on the training set and not only memorized the interactions.
~Table \ref{tab:avg_theoretical_max_sizebin_US_cables_city} , ~Table \ref{tab:avg_theoretical_max_sizebin_senate_bills} indicates the average theoretical maximum nodes to guess per hyperedge size (across folds) for CableGate City and senate-bills datasets, rest of the results on other datasets are presented in Appendix~\ref{appendix:hyperedge-prediction-results}.

\begin{figure}[h!]
    \centering
    \begin{tikzpicture}
    \begin{axis}[
        ybar,
        width=0.9\linewidth,
        height=0.55\linewidth,
        bar width=8pt,
        enlarge x limits=0.18,
        ymin=0,
        ymax=0.45, 
        ylabel={Average ratio},
        xlabel={Hyperedge Size},
        xtick=data,
        symbolic x coords={3--6,7--10,11--71},
        legend style={at={(0.5,1.02)},anchor=south,legend columns=-1},
        ymajorgrids=true,
        grid style={dashed},
        error bars/y dir=both,
        error bars/y explicit,
    ]

    \addplot+[draw=black, fill=blue!50] coordinates {
      (3--6, 0.1590) +- (0, 0.0080)
      (7--10, 0.1836) +- (0, 0.0023)
      (11--71, 0.2385) +- (0, 0.0038)
    };
    \addlegendentry{EDVW\_clique}

    \addplot+[draw=black, fill=orange!80] coordinates {
      (3--6, 0.1497) +- (0, 0.0078)
      (7--10, 0.1772) +- (0, 0.0020)
      (11--71, 0.2367) +- (0, 0.0036)
    };
    \addlegendentry{EDVW\_hyper}

    \addplot+[draw=black, fill=green!60] coordinates {
      (3--6, 0.1685) +- (0, 0.0104)
      (7--10, 0.2448) +- (0, 0.0072)
      (11--71, 0.3926) +- (0, 0.0056)
    };
    \addlegendentry{hyperwalk}

    \addplot+[draw=black, fill=red!40] coordinates {
      (3--6, 0.0093) +- (0, 0.0031)
      (7--10, 0.0142) +- (0, 0.0028)
      (11--71, 0.0295) +- (0, 0.0021)
    };
    \addlegendentry{random}

    \end{axis}
    \end{tikzpicture}
    \caption{US Cables City dataset performances by size bin for seen hyperedges (mean $\pm$ std across folds), $\alpha=0.5$.}
    \label{fig:TASK2-US-cables-city-seen-ratio}\end{figure}
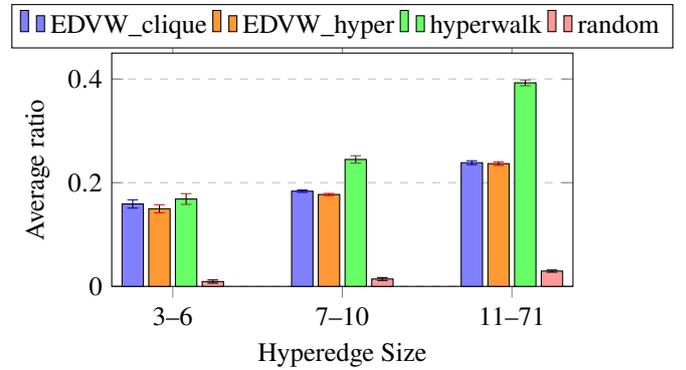


\paragraph{}
On the U.S. Cables City dataset, the EDVW hypergraph paired with the non-Markovian \textit{hyperwalk} consistently matches or outperforms all other \textit{Network Architecture + Random Walk} combinations across hyperedge sizes. Crucially, the performance gap between the non-Markovian \textit{hyperwalk} on the hypergraph and the other methods widens as hyperedge size increases.
These results show that the approach can predict consulate--embassy interactions that were never observed in the training set \(E^{T}\)---i.e., hyperedges \(\hat{e} \notin E^{T}\)---with especially strong gains under the \textit{hyperwalk}. This demonstrates that the EDVW hypergraph, when paired with a non-Markovian random walk, captures the underlying structure of diplomatic relationships and can infer new interactions.

On the senate-bills dataset, we observe the same trend, with the EDVW hypergraph combined with the non-Markovian random walk outperforming all other combos of \textit{Network Architecture + Random Walk} across all hyperedges sizes, and with higher performances on the novel interactions. However, the EDVW Hypergraph Markovian Walk actually outperforms the random walk on the clique graph on the largest hyperedges, showing that in some cases, the higher-order could be leverage by Markovian dynamics.
Those results show us that the hypergraphs framework could be irrelevant if not used with an adapted dynamics, even if the latter (i.e. EDVW Hypergraph Markovian Random Walk) is not equivalent to a dynamic on a projected graph \cite{neuhauser_multi-body_2019}.

\section{Discussion \& Conclusion}
In this work, we introduced a hypergraph pipeline with edge-dependent vertex weights (EDVW) to uncover higher-order interactions and applied it to U.S. diplomatic communications using the \textit{CableGate} corpus. We showed that this higher-order representation, when paired with non-Markovian dynamics, significantly outperformed traditional pairwise-graph baselines on fake-hyperedge detection and hyperedge prediction. 
Critically, the random-walk dynamics did not merely recover known ties; they also predicted previously real unobserved diplomatic interactions. We observed similar patterns on Email-Eu, Email-Enron, and Senate-Bills, although magnitudes varied; notably, Senate-Bills benefited most from the hypergraph formulation. 
In some settings, a Markovian random walk on the hypergraph also surpassed a walk on the clique-projected graph, indicating that higher-order structure could be exploited by Markovian dynamics as well. 
Methodologically, we provided a practical test for when hypergraph modeling was warranted by comparing combinations of \textit{Network Architecture + Random Walk} across self-supervised tasks on held-out data. 
Overall, we showed that the choice of random-walk dynamics was pivotal and, more generally, that dynamics should not be studied in isolation from network architecture. As a next step, the framework could be extended to temporal dynamics, including symmetry breaking via directed-hypergraph formulation.
\section*{Acknowledgements}
This work has benefited from insights provided in an interview with Alexa O’Brien, 
an investigative journalist and researcher who collaborated with WikiLeaks on the extensively documented Chelsea Manning court-martial. O’Brien has also conducted significant research and reporting on WikiLeaks’ publication of the United States diplomatic cables, 
which informed aspects of this study.

\normalsize
\bibliographystyle{unsrtnat}
\bibliography{references}

\clearpage
\thispagestyle{empty}
\mbox{}     
\section{Materials \& Methods Appendix}
\subsection{Cleaning and Validating the CableGate Dataset}
The objective here is to clean and validate the CableGate dataset, ensuring that it is ready for analysis. This includes removing duplicates, checking for inconsistencies, and validating the data against known sources.
\subsubsection{Cleaning Process}
Only include entities (sender and receiver) that are labeled as Embassies or Consulates regarding the wikileaks nomenclature \footnote{\url{https://wikileaks.org/plusd/about-fr/}, \url{https://wikileaks.org/plusd/about-to/}}. Attention, certain entity in the nomenclature are not explicitly labeled as Embassies or Consulates, such as cities name. To handle this, we verified that an embassy or consulate of US exists in the city (to ensure that it was not an isolated US Mission)
Summary of the constructed hypergraph: At the city/embassy-consulate level, the hypergraph contains 27,153 edges and 259 nodes. At the country level, it consists of 23,570 edges and 177 nodes. Note: No restrictions were applied based on the classification of the cables.
All cables have been gathered from the wikileaks website \footnote{https://wikileaks.org/plusd/cables/}, covering the period from 2000 to 2012.

\subsection{Formation of the Hypergraph on Common Datasets}
\label{appendix:common-datasets-formation}
\subsubsection{Email-Eu \& Email-Enron Datasets}

Similarly to the CableGate dataset, we constructed hypergraphs from the Email-Eu and Email-Enron datasets. In these datasets, each email is represented as a hyperedge connecting all recipients of the email (i.e., \{\textit{sender} $\cup$ \textit{receivers}\}). 
The vertex set consists of all email addresses involved in the communications.
The matrices \( R \) and \( W \) are defined as follows:
\begin{itemize}\setlength{\itemsep}{2pt}
    \item \textit{Homogeneous edge} (all known-party sponsors share one party): 
          \(R(e,v)=1\) for \(v\in e\); otherwise \(0\).
    \item \textit{Mixed edge} (both parties present): 
          \(R(e,v)=1+s_v\,\mathrm{sgn}(r_e)\,(1-|r_e|)\) for \(v\in e\); otherwise \(0\).
    \item \(W(v,e)=|e|\) if \(v\in e\); otherwise \(0\).
  \end{itemize}
  {\footnotesize
  \(e=\{v\in e:\ p(v)\in\{\mathrm{D},\mathrm{R}\}\}\),\quad
  \(r_e=\dfrac{D_e-R_e}{D_e+R_e}\) (set \(r_e=0\) if \(D_e+R_e=0\)),\quad
  \(s_v=\begin{cases}+1,&\text{Dem}\\-1,&\text{Rep}\end{cases}\).
  }

\noindent
Thus $R$ encodes edge–dependent (party–aware) incidence strengths, amplifying close majority–party sponsors on mixed edges and equalizing weights on homogeneous edges, while $W$ provides a per–edge vertex weight scaled by the number of known–party co-sponsors.
\subsection{Overview of Negative Sampling Procedures}

We implement three distinct negative sampling strategies to generate fake hyperedges \( E_f \) from the missing edge set \( E_m \). Each method differs in how it perturbs real edges while avoiding overlap with the original set \( E \).

\begin{itemize}[leftmargin=*] 
    \item \textbf{Alpha:} Replaces a fraction \( 1 - \alpha \) of nodes in each edge.
    \begin{itemize}[leftmargin=1.5em] 
        \item \textit{Control:} Continuous parameter \( \alpha \in [0, 1] \).
    \end{itemize}

    \item \textbf{$K$-Replacement:} Replaces exactly \( k \) nodes in each edge.
    \begin{itemize}[leftmargin=1.5em]
        \item \textit{Control:} Discrete number \( k < |e| \).
    \end{itemize}

    \item \textbf{Degree-Matched:} Generates fake edges with node degrees similar to those in \( E_m \).
    \begin{itemize}[leftmargin=1.5em]
        \item \textit{Control:} Degree-based matching via inverse indexing and binary search.
        \item \textit{Use case:} Appropriate for evaluations sensitive to degree distributions (Preserves structural properties of node degrees). Moreover, it makes the edge prediction task harder compared to the previous methods.
    \end{itemize}
\end{itemize}

\subsection{Scoring Method}
\label{appendix:scoring-method}
To compute an hyperedge score, the idea is to find a score that carries the information about the similarity between an arbitrary set of nodes. Moreover, the score should respects the condition: the higher the score is, the more probable that the edge exists. 
Moreover, the score should encapsulates the higher-order interactions, it should not rely on pairwise similarity between nodes. Those conditions lead us to chose the following scoring method: 

\begin{equation}
    S = \frac{1}{K}\sum_{k=1}^{K} P^{k}, \tag{2}
    \end{equation}
    where $K$ is the maximum length of the random walk.

\begin{equation}
\mathrm{JS}_{\pi}(p_1(x), p_2(x), \ldots, p_t(x)) 
= \sum_i \pi_i \sum_x p_i(x) \log_2 \frac{p_i(x)}{r(x)},
\end{equation}
Where $S$ is the similarity matrix, inspired from \cite{liu_link_2010}, and $P$ is the transition matrix of the random walk used.
$r(x) = \sum_i \pi_i p_i(x)$. The generalized Jensen--Shannon divergence (GJS) 
is invariant to any permutation of the distributions and is bounded above by $\log_2 t$. 
Therefore, the way we define the hyperedge score as (the LRW-GJS index):

\begin{equation}
S_e = 1 - \mathrm{JS}_{\pi_t}(S_{i_1}, \ldots, S_{i_t}) / \log_2 t,
\end{equation}
    
where $t = |e|$, $\pi_t = (1/t, \ldots, 1/t)$ and $e = \{ i_1, \ldots, i_t \}$.

\subsection{Hyperedge Prediction Algorithm}
Guessing nodes algorithm in \textit{Hyperedge Prediction Task}.\\

\begin{algorithm}[t!]
    \footnotesize
    \caption{Guess Neighbors with Refinement (cycle pivot)}
    \label{alg:guess-refine}
    \begin{algorithmic}[1]
    \Require Probes $E^{P}$; score list $S_{\text{steps}}$ (built on $E^{T}$ with $E=E^{T}\cup E^{P}$, $E^{T}\cap E^{P}=\emptyset$); node set $V$
    \Require Hyperparameters: $n_{\text{passes}}$, \texttt{early\_stop} (bool), \texttt{patience}$\ge1$; number of trials \texttt{trial}
    \Function{GreedyFill}{$\textit{preserved},\,\textit{must\_keep},\,n,\,S_{\text{steps}},\,V_{\text{pool}}$}
      \State Initialize \textit{chosen} with \textit{must\_keep}; set \textit{used} to $\textit{preserved}\cup\textit{must\_keep}$
      \For{$t=1$ to $n$}
        \State Let $S$ be $S_{\text{steps}}[\min(t, |S_{\text{steps}}|-1)]$
        \State Let $j^\star$ be the node in $V_{\text{pool}}\setminus \textit{used}$ maximizing $\mathrm{LRW\mbox{-}GJS}(\textit{preserved}\cup\textit{chosen}\cup\{j\};\,S)$
        \If{$j^\star$ does not exist} \textbf{break} \EndIf
        \State Append $j^\star$ to \textit{chosen} and add $j^\star$ to \textit{used}
      \EndFor
      \State \Return \textit{chosen}
    \EndFunction
    \Statex
    \For{$t=1$ \textbf{to} \texttt{trial}}
      \State Sample an incomplete $e_i\in E^{P}$; split into \textit{preserved} and \textit{target}
      \State Set $m$ to $|\textit{target}|$; set $V_{\text{pool}}$ to $V\setminus \textit{preserved}$; let $S_{\text{last}}$ be the last element of $S_{\text{steps}}$
      \State \textbf{Greedy fill:} set \textit{guessed} to \Call{GreedyFill}{$\textit{preserved},\,\emptyset,\,m,\,S_{\text{steps}},\,V_{\text{pool}}$}
      \State Set \textit{completed} to $\textit{preserved}\cup\textit{guessed}$; set \textit{best} to $\mathrm{LRW\mbox{-}GJS}(\textit{completed};\,S_{\text{last}})$; set $noImp$ to $0$
      \State \textbf{Refinement (cycle strategy):}
      \For{$p=1$ \textbf{to} $n_{\text{passes}}$}
        \If{\textit{guessed} is empty} \textbf{break} \EndIf
        \State Choose pivot $\pi$ as $\textit{guessed}[(p-1)\bmod |\textit{guessed}|]$ \Comment{cycle through guessed nodes}
        \State Set $n'$ to $\max(0,\,m-1)$
        \State Set \textit{new} to \Call{GreedyFill}{$\textit{preserved},\,\{\pi\},\,n',\,S_{\text{steps}},\,V_{\text{pool}}$}
        \State Set $s$ to $\mathrm{LRW\mbox{-}GJS}(\textit{preserved}\cup\textit{new};\,S_{\text{last}})$
        \If{$s$ strictly greater than \textit{best}}
          \State Update \textit{guessed} to \textit{new}; update \textit{completed} to $\textit{preserved}\cup\textit{new}$; update \textit{best} to $s$; set $noImp$ to $0$
        \Else
          \State Increment $noImp$ by $1$
          \If{\texttt{early\_stop} is true \textbf{and} $noImp$ at least \texttt{patience}} \textbf{break} \EndIf
        \EndIf
      \EndFor
    \EndFor
    \State \Return guessed sets
    \end{algorithmic}
    \vspace{-0.4em}
    \end{algorithm}
    
\clearpage
\onecolumn
\newpage
\subsection{Properties of Markov Chains}
\label{appendix:random-walk-properties}
\paragraph{}
\textit{Senate-Bills}: $N$ walks $= 10000$, $K_{max} = 120$. 
\textit{Email-EU}: $N$ walks $= 10000$, $K_{max} = 50$.
\textit{US-Cables-City}: $N$ walks $= 10000$, $K_{max} = 100$.
\textit{US-Cables-Country}: $N$ walks $= 10000$, $K_{max} = 100$.
\textit{Email-Enron}: $N$ walks $= 10000$, $K_{max} = 500$.
\textit{US-Cables-Clean-Ind}: $N$ walks $= 10000$, $K_{max} = 100$.
\begin{table}[H]
    \centering
    \caption{\textbf{Markov EDVW Hypergraph Walk:} Summary of Detailed Balance Checks Across Dataset }
    \label{tab:markov-random-walk-properties}
    \begin{tabular}{lcccccc}
    \toprule
    \textbf{Dataset} & \textbf{Reversible} & \textbf{Max} & \textbf{Mean Max} & \textbf{Total} & \textbf{Moderate} & \textbf{Severe} \\
                        & \textbf{folds (/10)} & \textbf{Violation} & \textbf{Violation} & \textbf{Violations} & \textbf{Violations} & \textbf{Violations} \\
    \midrule
    Senate-Bills            & 0 / 10 & $2.93 \times 10^{-5}$ & $2.62 \times 10^{-5}$ & 281494 & 271930 & 9564 \\
    US Cables City         & 0 / 10 & $ 1.42 \times 10^{-4}$ & $1.37 \times 10^{-4}$ & 87798 & 79796  & 8002 \\
    US Cables Country         & 0 / 10 & $ 4.02 \times 10^{-4}$ & $2.57 \times 10^{-4}$ & 190602 & 161134  & 29468 \\
    Email-Eu                & 0 / 10 & $9.15 \times 10^{-5}$ & $8.58 \times 10^{-5}$ & 247312 & 237358 & 9954 \\
    Email-Enron             & 0 / 10 & $8.57 \times 10^{-4}$ & $8.29 \times 10^{-4}$ & 41182  & 18844  & 22338 \\
    \bottomrule
    \end{tabular}
\end{table}
    
\begin{table}[H]
    \centering
    \caption{\textbf{Non-Markovian EDVW Hypergraph Walk:} Summary of Detailed Balance Checks Across Dataset}
    \label{tab:non-markov-random-walk-properties}
    \begin{tabular}{lcccccc}
    \toprule
    \textbf{Dataset} & \textbf{Reversible} & \textbf{Max} & \textbf{Mean Max} & \textbf{Total} & \textbf{Moderate} & \textbf{Severe} \\
                        & \textbf{folds (/10)} & \textbf{Violation} & \textbf{Violation} & \textbf{Violations} & \textbf{Violations} & \textbf{Violations} \\
    \midrule
    Senate-Bills            & 0 / 10 & $ 7.97 \times 10^{-5}$ & $6.24 \times 10^{-5}$ & 330116 & 289714 & 40402 \\
    US Cables Country         & 0 / 10 & $ 4.02 \times 10^{-4}$ & $2.57 \times 10^{-4}$ & 190602 & 161134  & 29468 \\
    US Cables City         & 0 / 10 & $ 3.48 \times 10^{-4}$ & $2.45 \times 10^{-4}$ & 232776 & 202944  & 29832 \\
    Email-Eu                & 0 / 10 & $ 1.07 \times 10^{-4}$ & $8.39 \times 10^{-5}$ & 332258 & 313836 & 18422 \\
    Email-Enron             & 0 / 10 & $ 7.85 \times 10^{-4}$ & $5.69 \times 10^{-4}$ & 45534  & 15870  & 29664 \\
    \bottomrule
    \end{tabular}
\end{table}

\begin{table}[H]
    \centering
    \caption{\textbf{EDVW Clique Graphs Walk:} Summary of Detailed Balance Checks Across Dataset }
    \label{tab:clique-random-walk-properties}
    \begin{tabular}{lcccccc}
    \toprule
    \textbf{Dataset} & \textbf{Reversible} & \textbf{Max} & \textbf{Mean Max} & \textbf{Total} & \textbf{Moderate} & \textbf{Severe} \\
                        & \textbf{folds (/10)} & \textbf{Violation} & \textbf{Violation} & \textbf{Violations} & \textbf{Violations} & \textbf{Violations} \\
    \midrule
    Senate-Bills            & 10 / 10 & $1.52 \times 10^{-18}$ & $7.78 \times 10^{-19}$ & 0 & 0 & 0 \\
    US Cables Country     & 10 / 10 & $9.26 \times 10^{-18}$ & $6.17 \times 10^{-18}$ & 0 & 0 & 0 \\
    US Cables City        & 10 / 10 & $1.65 \times 10^{-17}$ & $1.08 \times 10^{-17}$ & 0 & 0 & 0 \\
    Email-EU                & 10 / 10 & $6.40 \times 10^{-17}$ & $2.14 \times 10^{-17}$ & 0 & 0 & 0 \\
    Email-Enron             & 10 / 10 & $ 1.62 \times 10^{-17}$ & $9.25 \times 10^{-18}$ & 0  & 0  & 0 \\
    \bottomrule
    \end{tabular}
\end{table}
\clearpage
\twocolumn

\subsection{Results Fake Hyperedge Detection Task}
\label{appendix:common-datasets-results}
\subsubsection{US Cables Country Scale}
\label{appendix:country-scale-results}
In  Tables~\ref{tab:best-AUC-US-Cables-Country-TASK1-alpha0.5},~\ref{tab:best-AUC-US-Cables-Country-TASK1-k-replace-2},~\ref{tab:best-AUC-US-Cables-Country-TASK1-degree-d2}  are presented the best performances of the different combos \textit{Network Architecture + Random Walk} on the \textit{TASK 1} of detection of fake hyperedges in the US CableGate dataset at country level, using respectively the \textit{Alpha}, \textit{K-Replace} and \textit{Degree-Matching} sampling methods.


\begin{table}[!ht]
    \centering
    \caption{Country Scale: Best AUC per method and size bin (fake sampling method \textit{Alpha-Value} with $\alpha=0.5$), reported as mean $\pm$ std of the per-fold maxima. The AUC is the mean over the folds. In each fold and for all the size bins, the number of samples $n$ for AUC is $1000$.}
    \label{tab:best-AUC-US-Cables-Country-TASK1-alpha0.5}
    \begin{tabular}{l l c}
        \toprule
        Hyperedge & Method & Mean $\pm$ Std \\
        \midrule
        3–6 & EDVW\_clique & 0.6593 $\pm$ 0.0255 \\
        3–6 & EDVW\_hyper & 0.6601 $\pm$ 0.0239 \\
        3–6 & \textbf{hyperwalk} & \textbf{0.6979} $\pm$ \textbf{0.0214} \\
        \midrule
        \addlinespace[0.5ex]
        7–10 & EDVW\_clique & 0.6950 $\pm$ 0.0207 \\
        7–10 & EDVW\_hyper & 0.6980 $\pm$ 0.0143 \\
        7–10 & \textbf{hyperwalk} & \textbf{0.7183} $\pm$ \textbf{0.0138} \\
        \midrule
        \addlinespace[0.5ex]
        11–71 & EDVW\_clique & 0.6354 $\pm$ 0.0232 \\
        11–71 & EDVW\_hyper & 0.6320 $\pm$ 0.0130 \\
        11–71 & \textbf{hyperwalk} & \textbf{0.6631} $\pm$ \textbf{0.0067} \\
        \bottomrule
      \end{tabular}
    \end{table}

\begin{table}[!ht]
    \centering
    \caption{Country Scale: Best AUC per method and size bin (fake sampling method \textit{K-Replace} with $k=2$), reported as mean $\pm$ std of the per-fold maxima. The AUC is the mean over the folds. In each fold and for all the size bins, the number of samples $n$ for AUC is $1000$.}
    \label{tab:best-AUC-US-Cables-Country-TASK1-k-replace-2}
    \begin{tabular}{l l c}
      \toprule
      Hyperedge & Method & Mean $\pm$ Std \\
      \midrule
      3–6 & EDVW\_clique & 0.7600 $\pm$ 0.0168 \\
      3–6 & EDVW\_hyper & 0.7572 $\pm$ 0.0154 \\
      3–6 & \textbf{hyperwalk} & \textbf{0.7824} $\pm$ \textbf{0.0200} \\
      \midrule
      \addlinespace[0.5ex]
      7–10 & EDVW\_clique & 0.8206 $\pm$ 0.0113 \\
      7–10 & EDVW\_hyper & 0.8184 $\pm$ 0.0090 \\
      7–10 & \textbf{hyperwalk} & \textbf{0.8235} $\pm$ \textbf{0.0102} \\
      \midrule
      \addlinespace[0.5ex]
      11–\(\infty\) & EDVW\_clique & 0.7424 $\pm$ 0.0083 \\
      11–\(\infty\) & EDVW\_hyper & 0.7423 $\pm$ 0.0073 \\
      11–\(\infty\) & \textbf{hyperwalk} & \textbf{0.7467} $\pm$ \textbf{0.0076} \\
      \bottomrule
    \end{tabular}
  \end{table}


\begin{table}[!ht]
    \centering
    \caption{Country Scale: Best AUC per method and size bin (fake sampling method \textit{K Degree-Matched} with $k=2$), reported as mean $\pm$ std of the per-fold maxima. The AUC is the mean over the folds. In each fold and for all the size bins, the number of samples $n$ for AUC is $1000$.}
    \label{tab:best-AUC-US-Cables-Country-TASK1-degree-d2}
    \begin{tabular}{l l c}
      \toprule
      Hyperedge & Method & Mean $\pm$ Std \\
      \midrule
      3–6 & EDVW\_clique & 0.5193 $\pm$ 0.0128 \\
      3–6 & \textbf{EDVW\_hyper} & \textbf{0.5212} $\pm$ \textbf{0.0101} \\
      3–6 & hyperwalk & 0.5172 $\pm$ 0.0057 \\
      \midrule
      \addlinespace[0.5ex]
      7–10 & EDVW\_clique & 0.5171 $\pm$ 0.0061 \\
      7–10 & EDVW\_hyper & 0.5172 $\pm$ 0.0066 \\
      7–10 & \textbf{hyperwalk} & \textbf{0.5188} $\pm$ \textbf{0.0057} \\
      \midrule
      \addlinespace[0.5ex]
      11–71 & EDVW\_clique & 0.5147 $\pm$ 0.0040 \\
      11–71 & EDVW\_hyper & 0.5132 $\pm$ 0.0032 \\
      11–71 & \textbf{hyperwalk} & \textbf{0.5180} $\pm$ \textbf{0.0042} \\
      \bottomrule
    \end{tabular}
  \end{table}
  
\subsubsection{Email-Enron}
In Tables~\ref{tab:TASK_1_EMAIL_ENRON},~\ref{tab:best-AUC-email-enron-TASK1-k-replace-k2},~\ref{tab:best-AUC-email-enron-TASK1-degree-d2} are presented the performances of the different models and architectures on the \textit{Email-Enron} dataset.


\begin{table}[h!]
    \centering
    \caption{Best AUC per method and size bin (fake sampling method \textit{Alpha-Value} with $\alpha=0.5$) on email-enron dataset, reported as mean $\pm$ std of the per-fold maxima. AUC samples number $n=1000$ for range $3-6$, $n=169$ for range $7-10$ and $n=64$ for range $11-$\(\infty\).}
    \label{tab:TASK_1_EMAIL_ENRON}
    \begin{tabular}{l l c}
      \toprule
      Hyperedge & Method & Mean $\pm$ Std \\
      \midrule
      2--2  & EDVW\_clique   & 0.6822 $\pm$ 0.0486 \\
      2--2  & EDVW\_hyper    & 0.6762 $\pm$ 0.0444 \\
      2--2  & \textbf{hyperwalk} & \textbf{0.6790} $\pm$ \textbf{0.0464} \\
      \midrule
      \addlinespace[0.5ex]
      3--6  & EDVW\_clique   & 0.9012 $\pm$ 0.0374 \\
      3--6  & EDVW\_hyper    & 0.8963 $\pm$ 0.0383 \\
      3--6  & \textbf{hyperwalk} & \textbf{0.9239} $\pm$ \textbf{0.0349} \\
      \midrule
      \addlinespace[0.5ex]
      7--10 & EDVW\_clique   & 1.0000 $\pm$ 0.0000 \\
      7--10 & EDVW\_hyper    & 1.0000 $\pm$ 0.0000 \\
      7--10 & \textbf{hyperwalk} & \textbf{1.0000} $\pm$ \textbf{0.0000} \\
      \midrule
      \addlinespace[0.5ex]
      11–37 & EDVW\_clique   & 1.0000 $\pm$ 0.0000 \\
      11–37 & EDVW\_hyper    & 1.0000 $\pm$ 0.0000 \\
      11–37 & \textbf{hyperwalk} & \textbf{1.0000} $\pm$ \textbf{0.0000} \\
      \bottomrule
    \end{tabular}
  \end{table}

\begin{table}[!ht]
    \centering
    \caption{Best AUC per method and size bin (fake sampling method \textit{K-Replace} with $k=2$) on email-enron dataset, reported as mean $\pm$ std of the per-fold maxima. AUC samples number $n=1000$ for range $3-6$, $n=169$ for range $7-10$ and $n=64$ for range $11-$\(\infty\).}
    \label{tab:best-AUC-email-enron-TASK1-k-replace-k2}
    \begin{tabular}{l l c}
      \toprule
      Hyperedge & Method & Mean $\pm$ Std \\
      \midrule
      3–6 & EDVW\_clique & 0.9154 $\pm$ 0.0241 \\
      3–6 & EDVW\_hyper & 0.9083 $\pm$ 0.0253 \\
      3–6 & \textbf{hyperwalk} & \textbf{0.9358} $\pm$ \textbf{0.0235} \\
      \midrule
      \addlinespace[0.5ex]
      7–10 & EDVW\_clique & 0.9620 $\pm$ 0.0471 \\
      7–10 & EDVW\_hyper & 0.9472 $\pm$ 0.0476 \\
      7–10 & \textbf{hyperwalk} & \textbf{0.9922} $\pm$ \textbf{0.0187} \\
      \midrule
      \addlinespace[0.5ex]
      11–37 & EDVW\_clique & 0.9611 $\pm$ 0.0571 \\
      11–37 & EDVW\_hyper & 0.9769 $\pm$ 0.0408 \\
      11–37 & \textbf{hyperwalk} & \textbf{0.9960} $\pm$ \textbf{0.0126} \\
      \bottomrule
    \end{tabular}
  \end{table}

\begin{table}[!ht]
    \centering
    \caption{Best AUC per method and size bin (fake sampling method \textit{K Degree-Matched} with $k=2$) on email-enron dataset, reported as mean $\pm$ std of the per-fold maxima. AUC samples number $n=1000$ for range $3-6$, $n=169$ for range $7-10$ and $n=64$ for range $11-$\(\infty\).}
    \label{tab:best-AUC-email-enron-TASK1-degree-d2}
    \begin{tabular}{l l c}
      \toprule
      Hyperedge & Method & Mean $\pm$ Std \\
      \midrule
      3–6 & EDVW\_clique & 0.7768 $\pm$ 0.0434 \\
      3–6 & EDVW\_hyper & 0.7759 $\pm$ 0.0422 \\
      3–6 & \textbf{hyperwalk} & \textbf{0.7981} $\pm$ \textbf{0.0332} \\
      \midrule
      \addlinespace[0.5ex]
      7–10 & EDVW\_clique & 0.8304 $\pm$ 0.0897 \\
      7–10 & EDVW\_hyper & 0.8352 $\pm$ 0.0797 \\
      7–10 & \textbf{hyperwalk} & \textbf{0.8872} $\pm$ \textbf{0.0831} \\
      \midrule
      \addlinespace[0.5ex]
      11–37 & EDVW\_clique & 0.8762 $\pm$ 0.1135 \\
      11–37 & EDVW\_hyper & 0.8717 $\pm$ 0.1158 \\
      11–37 & \textbf{hyperwalk} & \textbf{0.9146} $\pm$ \textbf{0.0948} \\
      \bottomrule
    \end{tabular}
  \end{table}

\subsubsection{Email-Eu}
\label{appendix:email-eu-results}
In Tables~\ref{tab:best-AUC-email-eu-TASK1-alpha-0.5},~\ref{tab:best-AUC-email-eu-TASK1-degree-d2},~\ref{tab:best-AUC-email-eu-TASK1-k-replace-k2} are presented the performances of the different models and architectures on the \textit{Email-Enron} dataset.

\begin{table}[!ht]
  \centering
  \caption{Best AUC per method and size bin (fake sampling method \textit{Alpha-Value} with $\alpha=0.5$) on email-eu dataset, reported as mean $\pm$ std of the per-fold maxima. AUC samples number $n=1000$.}
  \label{tab:best-AUC-email-eu-TASK1-alpha-0.5}
  \begin{tabular}{l l c}
    \toprule
    Hyperedge & Method & Mean $\pm$ Std \\
    \midrule
    3–6 & EDVW\_clique & 0.9177 $\pm$ 0.0084 \\
    3–6 & EDVW\_hyper & 0.9148 $\pm$ 0.0066 \\
    3–6 & \textbf{hyperwalk} & \textbf{0.9213} $\pm$ \textbf{0.0084} \\
    \midrule
    \addlinespace[0.5ex]
    7–10 & \textbf{EDVW\_clique} & \textbf{0.9886} $\pm$ \textbf{0.0048} \\
    7–10 & EDVW\_hyper & 0.9879 $\pm$ 0.0046 \\
    7–10 & hyperwalk & 0.9865 $\pm$ 0.0062 \\
    \midrule
    \addlinespace[0.5ex]
    11–40 & EDVW\_clique & 0.9878 $\pm$ 0.0091 \\
    11–40 & \textbf{EDVW\_hyper} & \textbf{0.9886} $\pm$ \textbf{0.0079} \\
    11–40 & hyperwalk & 0.9848 $\pm$ 0.0088 \\
    \bottomrule
  \end{tabular}
\end{table}

\begin{table}[!ht]
    \centering
    \caption{Best AUC per method and size bin (fake sampling method \textit{K-Replace} with $K=2$) on email-eu dataset, reported as mean $\pm$ std of the per-fold maxima. The AUC is the mean over folds of the per-fold max AUC. For all size bins and folds, the number of AUC samples is $n=1000$.}
    \label{tab:best-AUC-email-eu-TASK1-k-replace-k2}
    \begin{tabular}{l l c}
      \toprule
      Hyperedge & Method & Mean $\pm$ Std \\
      \midrule
      3–6 & EDVW\_clique & 0.9396 $\pm$ 0.0054 \\
      3–6 & EDVW\_hyper & 0.9385 $\pm$ 0.0041 \\
      3–6 & \textbf{hyperwalk} & \textbf{0.9398} $\pm$ \textbf{0.0043} \\
      \midrule
      \addlinespace[0.5ex]
      7–10 & \textbf{EDVW\_clique} & \textbf{0.9258} $\pm$ \textbf{0.0105} \\
      7–10 & EDVW\_hyper & 0.9225 $\pm$ 0.0100 \\
      7–10 & hyperwalk & 0.9144 $\pm$ 0.0132 \\
      \midrule
      \addlinespace[0.5ex]
      11–40 & \textbf{EDVW\_clique} & \textbf{0.8125} $\pm$ \textbf{0.0121} \\
      11–40 & EDVW\_hyper & 0.8122 $\pm$ 0.0143 \\
      11–40 & hyperwalk & 0.7777 $\pm$ 0.0161 \\
      \bottomrule
    \end{tabular}
  \end{table}

\begin{table}[!ht]
    \centering
    \caption{Best AUC per method and size bin (fake sampling method \textit{K Degree-Matched} with $K=2$) on email-eu dataset, reported as mean $\pm$ std of the per-fold maxima. The AUC is the mean over folds of the per-fold max AUC. For all size bins and folds, the number of AUC samples is $n=1000$.}
    \label{tab:best-AUC-email-eu-TASK1-degree-d2}
    \begin{tabular}{l l c}
      \toprule
      Hyperedge & Method & Mean $\pm$ Std \\
      \midrule
      3–6 & EDVW\_clique & 0.7022 $\pm$ 0.0116 \\
      3–6 & EDVW\_hyper & 0.7017 $\pm$ 0.0127 \\
      3–6 & \textbf{hyperwalk} & \textbf{0.7052} $\pm$ \textbf{0.0131} \\
      \midrule
      \addlinespace[0.5ex]
      7–10 & \textbf{EDVW\_clique} & \textbf{0.7002} $\pm$ \textbf{0.0241} \\
      7–10 & EDVW\_hyper & 0.6986 $\pm$ 0.0244 \\
      7–10 & hyperwalk & 0.6951 $\pm$ 0.0202 \\
      \midrule
      \addlinespace[0.5ex]
      11–40 & \textbf{EDVW\_clique} & \textbf{0.6790} $\pm$ \textbf{0.0185} \\
      11–40 & EDVW\_hyper & 0.6755 $\pm$ 0.0181 \\
      11–40 & hyperwalk & 0.6569 $\pm$ 0.0205 \\
      \bottomrule
    \end{tabular}
  \end{table}

\subsubsection{Senate-bills}
In Tables~\ref{tab:best-AUC-senate-bills-TASK1-alpha-0.5},~\ref{tab:best-AUC-senate-bills-TASK1-degree-d2} are presented the performances of the different models and architectures on the \textit{Senate-Bills} dataset.

\begin{table}[!ht]
    \centering
    \caption{Best AUC per method and size bin (fake sampling method \textit{Alpha} with $\alpha=0.5$) on senate-bills dataset, reported as mean $\pm$ std of the per-fold maxima. The AUC is the mean over folds of the per-fold max AUC. For all size bins and folds, the number of AUC samples is $n=1000$.}
    \label{tab:best-AUC-senate-bills-TASK1-alpha-0.5}
    \begin{tabular}{l l c}
      \toprule
      Hyperedge & Method & Mean $\pm$ Std \\
      \midrule
      3–6 & EDVW\_clique & 0.8391 $\pm$ 0.0556 \\
      3–6 & EDVW\_hyper & 0.8440 $\pm$ 0.0566 \\
      3–6 & \textbf{hyperwalk} & \textbf{0.8892} $\pm$ \textbf{0.0583} \\
      \midrule
      \addlinespace[0.5ex]
      7–10 & EDVW\_clique & 0.9394 $\pm$ 0.0146 \\
      7–10 & EDVW\_hyper & 0.9408 $\pm$ 0.0219 \\
      7–10 & \textbf{hyperwalk} & \textbf{0.9669} $\pm$ \textbf{0.0078} \\
      \midrule
      \addlinespace[0.5ex]
      11–99 & EDVW\_clique & 0.9782 $\pm$ 0.0076 \\
      11–99 & EDVW\_hyper & 0.9840 $\pm$ 0.0073 \\
      11–99 & \textbf{hyperwalk} & \textbf{0.9988} $\pm$ \textbf{0.0012} \\
      \bottomrule
    \end{tabular}
  \end{table}

\begin{table}[!ht]
    \centering
    \caption{Best AUC per method and size bin (fake sampling method \textit{K Degree-Matched} with $K=2$) on senate-bills dataset, reported as mean $\pm$ std of the per-fold maxima. The AUC is the mean over folds of the per-fold max AUC. For all size bins and folds, the number of AUC samples is $n=1000$.}
    \label{tab:best-AUC-senate-bills-TASK1-degree-d2}
    \begin{tabular}{l l c}
      \toprule
      Hyperedge & Method & Mean $\pm$ Std \\
      \midrule
      3–6 & EDVW\_clique & 0.5753 $\pm$ 0.0211 \\
      3–6 & EDVW\_hyper & 0.5816 $\pm$ 0.0218 \\
      3–6 & \textbf{hyperwalk} & \textbf{0.5967} $\pm$ \textbf{0.0221} \\
      \midrule
      \addlinespace[0.5ex]
      7–10 & EDVW\_clique & 0.5339 $\pm$ 0.0098 \\
      7–10 & EDVW\_hyper & 0.5345 $\pm$ 0.0128 \\
      7–10 & \textbf{hyperwalk} & \textbf{0.5425} $\pm$ \textbf{0.0169} \\
      \midrule
      \addlinespace[0.5ex]
      11–99 & EDVW\_clique & 0.5208 $\pm$ 0.0024 \\
      11–99 & EDVW\_hyper & 0.5241 $\pm$ 0.0047 \\
      11–99 & \textbf{hyperwalk} & \textbf{0.5261} $\pm$ \textbf{0.0032} \\
      \bottomrule
    \end{tabular}
  \end{table}

\subsection{Intruder Detection Subtask}

\label{appendix:regression-method}
To perform the regression task, we used a simple exponential model to fit the data. The model is defined as follows:
\begin{equation}
    y = a \cdot e^{b \cdot x} + c
\end{equation}
Where: $y$ is the similarity score gap between true and fake hyperedges, $x$ is the number of steps $k$ used to compute the similarity matrix , and $a$, $b$, and $c$ are parameters to be estimated.
Important note, to fully capture the evolution of the gap score, the similarity matrix $S_{k}$ is computed as $S_{k} = P^{k}$, where $P$ is the transition matrix of the random walk used.
This version differs from the one used in the fake hyperedge detection task, where the similarity matrix is computed as $S = \frac{1}{K}\sum_{k=1}^{K} P^{k}$, in order here to fully capture the evolution of the gap score over $k$ without keeping contribution of local random walks.

\subsection{Results Hyperedge Prediction Task}
\label{appendix:hyperedge-prediction-results}
\label{appendix:hyperedge-prediction-results-US-cables-city}
\subsubsection{US Cables City Scale}
\begin{table}[!ht]
    \centering
    \caption{Average theoretical maximum nodes to guess per size bin (across folds).}
    \label{tab:avg_theoretical_max_sizebin_US_cables_city}
    \begin{tabular}{lrrr}
    \toprule
    Size bin & Overall & Seen & Novel \\
    \midrule
    3--6        &  5,116 &  4,788 &   328 \\
    7--10       & 18,057 & 14,745 & 3,312 \\
    11--$\infty$ & 55,638 & 23,384 & 32,255 \\
    \bottomrule
    \end{tabular}
\end{table}

\subsubsection{US Cables Country Scale}
\begin{figure}[h!]
    \centering
    \begin{tikzpicture}
    \begin{axis}[
        ybar,
        width=0.9\linewidth,
        height=0.55\linewidth,
        bar width=8pt,
        enlarge x limits=0.18,
        ymin=0,
        ymax=0.45, 
        ylabel={Average ratio},
        xlabel={Hyperedge Size},
        xtick=data,
        symbolic x coords={3--6,7--10,11--71},
        legend style={at={(0.5,1.02)},anchor=south,legend columns=-1},
        ymajorgrids=true,
        grid style={dashed},
        error bars/y dir=both,
        error bars/y explicit,
    ]

    \addplot+[draw=black, fill=blue!50] coordinates {
      (3--6, 0.1715) +- (0, 0.0103)
      (7--10, 0.2018) +- (0, 0.0068)
      (11--71, 0.2529) +- (0, 0.0064)
    };
    \addlegendentry{EDVW\_clique}

    \addplot+[draw=black, fill=orange!80] coordinates {
      (3--6, 0.1664) +- (0, 0.0106)
      (7--10, 0.1955) +- (0, 0.0072)
      (11--71, 0.2480) +- (0, 0.0068)
    };
    \addlegendentry{EDVW\_hyper}

    \addplot+[draw=black, fill=green!60] coordinates {
      (3--6, 0.1786) +- (0, 0.0076)
      (7--10, 0.2660) +- (0, 0.0098)
      (11--71, 0.4199) +- (0, 0.0068)
    };
    \addlegendentry{hyperwalk}

    \addplot+[draw=black, fill=red!40] coordinates {
      (3--6, 0.0164) +- (0, 0.0016)
      (7--10, 0.0256) +- (0, 0.0014)
      (11--71, 0.0702) +- (0, 0.0037)
    };
    \addlegendentry{random}

    \end{axis}
    \end{tikzpicture}
    \caption{US Cables Country dataset performances by size bin for overall hyperedges (mean $\pm$ std across folds), $\alpha=0.5$.}
    \label{fig:TASK2-US-cables-country-overall-ratio}\end{figure}
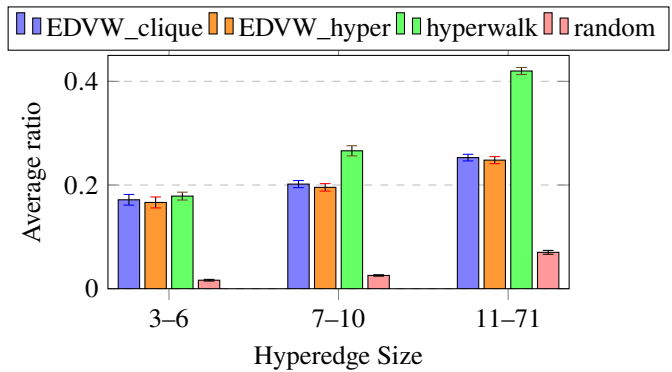

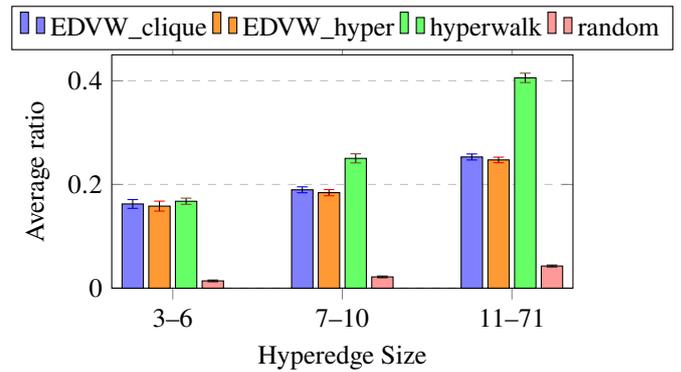
\begin{figure}[h!]
    \centering
    \begin{tikzpicture}
    \begin{axis}[
        ybar,
        width=0.9\linewidth,
        height=0.55\linewidth,
        bar width=8pt,
        enlarge x limits=0.18,
        ymin=0,
        ymax=0.45, 
        ylabel={Average ratio},
        xlabel={Hyperedge Size},
        xtick=data,
        symbolic x coords={3--6,7--10,11--71},
        legend style={at={(0.5,1.02)},anchor=south,legend columns=-1},
        ymajorgrids=true,
        grid style={dashed},
        error bars/y dir=both,
        error bars/y explicit,
    ]

    \addplot+[draw=black, fill=blue!50] coordinates {
      (3--6, 0.1624) +- (0, 0.0085)
      (7--10, 0.1897) +- (0, 0.0057)
      (11--71, 0.2530) +- (0, 0.0058)
    };
    \addlegendentry{EDVW\_clique}

    \addplot+[draw=black, fill=orange!80] coordinates {
      (3--6, 0.1582) +- (0, 0.0096)
      (7--10, 0.1842) +- (0, 0.0060)
      (11--71, 0.2473) +- (0, 0.0053)
    };
    \addlegendentry{EDVW\_hyper}

    \addplot+[draw=black, fill=green!60] coordinates {
      (3--6, 0.1676) +- (0, 0.0060)
      (7--10, 0.2503) +- (0, 0.0088)
      (11--71, 0.4056) +- (0, 0.0090)
    };
    \addlegendentry{hyperwalk}

    \addplot+[draw=black, fill=red!40] coordinates {
      (3--6, 0.0140) +- (0, 0.0016)
      (7--10, 0.0217) +- (0, 0.0016)
      (11--71, 0.0427) +- (0, 0.0018)
    };
    \addlegendentry{random}

    \end{axis}
    \end{tikzpicture}
    \caption{US Cables Country dataset performances by size bin for seen hyperedges (mean $\pm$ std across folds), $\alpha=0.5$.}
    \label{fig:TASK2-US-cables-country-seen-ratio}\end{figure}

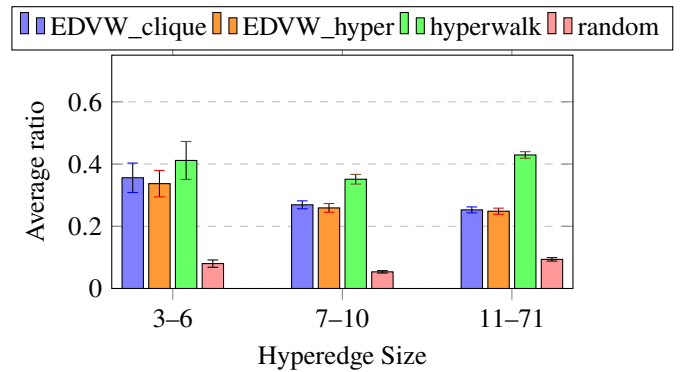
\begin{figure}[h!]
    \centering
    \begin{tikzpicture}
    \begin{axis}[
        ybar,
        width=0.9\linewidth,
        height=0.55\linewidth,
        bar width=8pt,
        enlarge x limits=0.18,
        ymin=0,
        ymax=0.75, 
        ylabel={Average ratio},
        xlabel={Hyperedge Size},
        xtick=data,
        symbolic x coords={3--6,7--10,11--71},
        legend style={at={(0.5,1.02)},anchor=south,legend columns=-1},
        ymajorgrids=true,
        grid style={dashed},
        error bars/y dir=both,
        error bars/y explicit,
    ]

    \addplot+[draw=black, fill=blue!50] coordinates {
      (3--6, 0.3559) +- (0, 0.0473)
      (7--10, 0.2690) +- (0, 0.0127)
      (11--71, 0.2527) +- (0, 0.0095)
    };
    \addlegendentry{EDVW\_clique}

    \addplot+[draw=black, fill=orange!80] coordinates {
      (3--6, 0.3370) +- (0, 0.0426)
      (7--10, 0.2590) +- (0, 0.0140)
      (11--71, 0.2484) +- (0, 0.0099)
    };
    \addlegendentry{EDVW\_hyper}

    \addplot+[draw=black, fill=green!60] coordinates {
      (3--6, 0.4116) +- (0, 0.0610)
      (7--10, 0.3512) +- (0, 0.0156)
      (11--71, 0.4292) +- (0, 0.0106)
    };
    \addlegendentry{hyperwalk}

    \addplot+[draw=black, fill=red!40] coordinates {
      (3--6, 0.0800) +- (0, 0.0117)
      (7--10, 0.0535) +- (0, 0.0039)
      (11--71, 0.0936) +- (0, 0.0055)
    };
    \addlegendentry{random}

    \end{axis}
    \end{tikzpicture}
    \caption{US Cables Country dataset performances by size bin for novel hyperedges (mean $\pm$ std across folds), $\alpha=0.5$.}
    \label{fig:TASK2-US-cables-country-novel-ratio}\end{figure}

Here are presented the performances on the \textit{US Cables Country Scale} dataset of the different combos \textit{Network Architecture + Random Walk} on the \textit{TASK 2} of hyperedge prediction.
In figure ~\ref{fig:TASK2-US-cables-country-overall-ratio} the overall ratio of corrects guessed nodes over the theoretical amount, in ~\ref{fig:TASK2-US-cables-country-novel-ratio}, the ratio of the \textit{novel} interactions (i.e. set \textit{preserved + correctly guessed} that is not a subset of any hyperedge in $E^{T}$)
and finally ~\ref{fig:TASK2-US-cables-country-seen-ratio} of the seen interactions (i.e. it exists at least one identical set \textit{preserved + correctly guessed} that is a subset of a hyperedge in $E^{T}$).


\begin{table}[htbp]
    \centering
    \caption{Average theoretical maximum nodes to guess per size bin (across folds) for US Cables Country dataset.}
    \label{tab:avg_theoretical_max_sizebin_us_cables_country}
    \begin{tabular}{lccc}
    \toprule
    Size bin & Overall & Seen & Novel \\
    \midrule
    3--6         & 16,793 & 16,074 & 719 \\
    7--10        & 53,990 & 46,128 & 7,862 \\
    11--$\infty$ & 162,249 & 69,116 & 93,134 \\
    \bottomrule
    \end{tabular}
\end{table}

\subsubsection{Email-Enron}
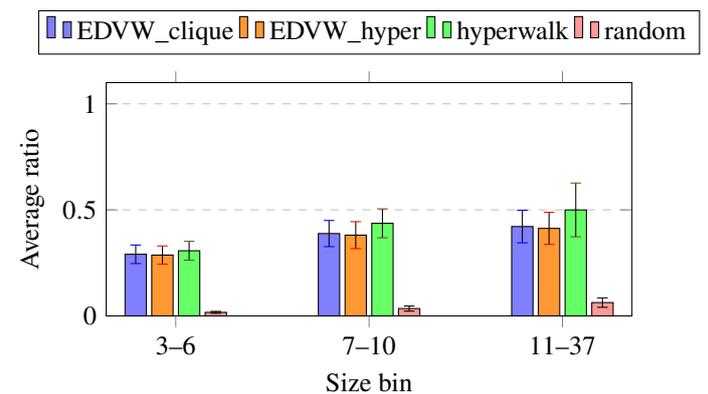
\begin{figure}[h!]
    \centering
    \begin{tikzpicture}
    \begin{axis}[
        ybar,
        width=\linewidth,
        height=0.55\linewidth,
        bar width=8pt,
        enlarge x limits=0.18,
        ymin=0,
        ymax=1.10,
        ylabel={Average ratio},
        xlabel={Size bin},
        xtick=data,
        symbolic x coords={2--2,3--6,7--10,11--37},
        legend style={at={(0.5,1.12)},anchor=south,legend columns=-1},
        ymajorgrids=true,
        grid style={dashed},
        error bars/y dir=both,
        error bars/y explicit,
    ]
    
    \addplot+[draw=black, fill=blue!50] coordinates {
        (3--6, 0.2906) +- (0, 0.0437)
        (7--10, 0.3883) +- (0, 0.0616)
        (11--37, 0.4214) +- (0, 0.0769)
    };
    \addlegendentry{EDVW\_clique}
    
    \addplot+[draw=black, fill=orange!80] coordinates {
        (3--6, 0.2868) +- (0, 0.0432)
        (7--10, 0.3808) +- (0, 0.0636)
        (11--37, 0.4128) +- (0, 0.0757)
    };
    \addlegendentry{EDVW\_hyper}
    
    \addplot+[draw=black, fill=green!60] coordinates {
        (3--6, 0.3072) +- (0, 0.0446)
        (7--10, 0.4366) +- (0, 0.0681)
        (11--37, 0.4997) +- (0, 0.1268)
    };
    \addlegendentry{hyperwalk}
    
    \addplot+[draw=black, fill=red!40] coordinates {
        (3--6, 0.0167) +- (0, 0.0045)
        (7--10, 0.0346) +- (0, 0.0121)
        (11--37, 0.0628) +- (0, 0.0223)
    };
    \addlegendentry{random}
    
    \end{axis}
    \end{tikzpicture}
    \caption{Email-Enron dataset performances by size bin for overall hyperedges (mean $\pm$ std across folds), $\alpha=0.5$.}
    \label{fig:TASK2-email-enron-overall-ratio}
  \end{figure}

Here are presented the performances on the \textit{Email-Enron} dataset of the different combos \textit{Network Architecture + Random Walk} on the \textit{TASK 2} of hyperedge prediction.
In figure ~\ref{fig:TASK2-email-enron-overall-ratio} the overall ratio of corrects guessed nodes over the theoretical amount, in ~\ref{fig:TASK2-email-enron-novel-ratio}, the ratio of the \textit{novel} interactions (i.e set \textit{preserved + correctly guessed} that is not a subset of any hyperedge in $E^{T}$)
and finally ~\ref{fig:TASK2-email-enron-seen-ratio} of the seen interactions (i.e. it exists at least one identical set \textit{preserved + correctly guessed} that is a subset of a hyperedge in $E^{T}$).


\begin{figure}[h!]
    \centering
    \begin{tikzpicture}
    \begin{axis}[
        ybar,
        width=\linewidth,
        height=0.55\linewidth,
        bar width=8pt,
        enlarge x limits=0.18,
        ymin=0,
        ymax=1.10,
        ylabel={Average ratio},
        xlabel={Size bin},
        xtick=data,
        symbolic x coords={2--2,3--6,7--10,11--37},
        legend style={at={(0.5,1.12)},anchor=south,legend columns=-1},
        ymajorgrids=true,
        grid style={dashed},
        error bars/y dir=both,
        error bars/y explicit,
    ]
    
    \addplot+[draw=black, fill=blue!50] coordinates {
        (3--6, 0.2346) +- (0, 0.0309)
        (7--10, 0.3471) +- (0, 0.0728)
        (11--37, 0.5046) +- (0, 0.0867)
    };
    \addlegendentry{EDVW\_clique}
    
    \addplot+[draw=black, fill=orange!80] coordinates {
        (3--6, 0.2338) +- (0, 0.0321)
        (7--10, 0.3333) +- (0, 0.0890)
        (11--37, 0.4767) +- (0, 0.0961)
    };
    \addlegendentry{EDVW\_hyper}
    
    \addplot+[draw=black, fill=green!60] coordinates {
        (3--6, 0.2374) +- (0, 0.0338)
        (7--10, 0.4191) +- (0, 0.1044)
        (11--37, 0.5843) +- (0, 0.1370)
    };
    \addlegendentry{hyperwalk}
    
    \addplot+[draw=black, fill=red!40] coordinates {
        (3--6, 0.0114) +- (0, 0.0028)
        (7--10, 0.0246) +- (0, 0.0110)
        (11--37, 0.0488) +- (0, 0.0340)
    };
    \addlegendentry{random}
    
    \end{axis}
    \end{tikzpicture}
    \caption{Email-Enron dataset performances by size bin for seen hyperedges (mean $\pm$ std across folds), $\alpha=0.5$.}
    \label{fig:TASK2-email-enron-seen-ratio}
  \end{figure}

\begin{figure}[h!]
    \centering
    \begin{tikzpicture}
    \begin{axis}[
        ybar,
        width=\linewidth,
        height=0.55\linewidth,
        bar width=8pt,
        enlarge x limits=0.18,
        ymin=0,
        ymax=1.10,
        ylabel={Average ratio},
        xlabel={Size bin},
        xtick=data,
        symbolic x coords={3--6,7--10,11--37},
        legend style={at={(0.5,1.12)},anchor=south,legend columns=-1},
        ymajorgrids=true,
        grid style={dashed},
        error bars/y dir=both,
        error bars/y explicit,
    ]
    
    \addplot+[draw=black, fill=blue!50] coordinates {
        (3--6, 0.4969) +- (0, 0.0697)
        (7--10, 0.4105) +- (0, 0.0671)
        (11--37, 0.4123) +- (0, 0.0852)
    };
    \addlegendentry{EDVW\_clique}
    
    \addplot+[draw=black, fill=orange!80] coordinates {
        (3--6, 0.4861) +- (0, 0.0633)
        (7--10, 0.4061) +- (0, 0.0623)
        (11--37, 0.4048) +- (0, 0.0812)
    };
    \addlegendentry{EDVW\_hyper}
    
    \addplot+[draw=black, fill=green!60] coordinates {
        (3--6, 0.5580) +- (0, 0.0713)
        (7--10, 0.4539) +- (0, 0.0588)
        (11--37, 0.4909) +- (0, 0.1364)
    };
    \addlegendentry{hyperwalk}
    
    \addplot+[draw=black, fill=red!40] coordinates {
        (3--6, 0.0720) +- (0, 0.0437)
        (7--10, 0.0471) +- (0, 0.0173)
        (11--37, 0.0642) +- (0, 0.0234)
    };
    \addlegendentry{random}
    
    \end{axis}
    \end{tikzpicture}
    \caption{Email-Enron dataset performances by size bin for novel hyperedges (mean $\pm$ std across folds), $\alpha=0.5$.}
    \label{fig:TASK2-email-enron-novel-ratio}
  \end{figure}

\begin{table}[htbp]
    \centering
    \caption{Average theoretical maximum nodes to guess per size bin (across folds) for Email-Enron dataset.}
    \label{tab:avg_theoretical_max_sizebin}
    \begin{tabular}{lrrr}
    \toprule
    Size bin & Overall & Seen & Novel \\
    \midrule
    2--2        &   235 &   233 &     2 \\
    3--6        &   503 &   418 &    85 \\
    7--10       &   322 &   137 &   185 \\
    11--$\infty$ &   338 &    38 &   299 \\
    \bottomrule
    \end{tabular}
\end{table}

\subsubsection{Email-EU}
Here are presented the performances on the \textit{Email-Eu} dataset of the different combos \textit{Network Architecture + Random Walk} on the \textit{TASK 2} of hyperedge prediction.
In figure ~\ref{fig:TASK2-email-eu-overall-ratio} the overall ratio of corrects guessed nodes over the theoretical amount, in ~\ref{fig:TASK2-email-eu-novel-ratio}, the ratio of the \textit{novel} interactions (i.e set \textit{preserved + correctly guessed} that is not a subset of any hyperedge in $E^{T}$)
and finally ~\ref{fig:TASK2-email-eu-seen-ratio} of the seen interactions (i.e. it exists at least one identical set \textit{preserved + correctly guessed} that is a subset of a hyperedge in $E^{T}$).
~Table \ref{tab:avg_theoretical_max_sizebin_email_eu} indicates the average theoretical maximum nodes to guess per size bin (across folds).

\begin{figure}[h!]
    \centering
    \begin{tikzpicture}
    \begin{axis}[
        ybar,
        width=\linewidth,
        height=0.55\linewidth,
        bar width=8pt,
        enlarge x limits=0.18,
        ymin=0,
        ymax=0.60,
        ylabel={Average ratio},
        xlabel={Size bin},
        xtick=data,
        symbolic x coords={3--6,7--10,11--40},
        legend style={at={(0.5,1.12)},anchor=south,legend columns=-1},
        ymajorgrids=true,
        grid style={dashed},
        error bars/y dir=both,
        error bars/y explicit,
    ]
    
    \addplot+[draw=black, fill=blue!50] coordinates {
        (3--6, 0.1573) +- (0, 0.0062)
        (7--10, 0.2806) +- (0, 0.0109)
        (11--40, 0.3755) +- (0, 0.0253)
    };
    \addlegendentry{EDVW\_clique}
    
    \addplot+[draw=black, fill=orange!80] coordinates {
        (3--6, 0.1536) +- (0, 0.0064)
        (7--10, 0.2766) +- (0, 0.0099)
        (11--40, 0.3751) +- (0, 0.0249)
    };
    \addlegendentry{EDVW\_hyper}
    
    \addplot+[draw=black, fill=green!60] coordinates {
        (3--6, 0.1638) +- (0, 0.0083)
        (7--10, 0.3212) +- (0, 0.0141)
        (11--40, 0.4506) +- (0, 0.0214)
    };
    \addlegendentry{hyperwalk}
    
    \addplot+[draw=black, fill=red!40] coordinates {
        (3--6, 0.0020) +- (0, 0.0008)
        (7--10, 0.0051) +- (0, 0.0009)
        (11--40, 0.0111) +- (0, 0.0019)
    };
    \addlegendentry{random}
    
    \end{axis}
    \end{tikzpicture}
    \caption{Email-EU dataset performances by size bin for overall hyperedges (mean $\pm$ std across folds), $\alpha=0.5$.}
    \label{fig:TASK2-email-eu-overall-ratio}
  \end{figure}
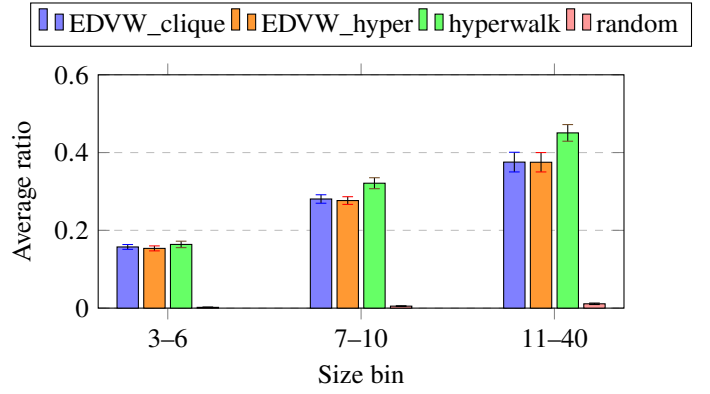


\begin{figure}[h!]
    \centering
    \begin{tikzpicture}
    \begin{axis}[
        ybar,
        width=\linewidth,
        height=0.55\linewidth,
        bar width=8pt,
        enlarge x limits=0.18,
        ymin=0,
        ymax=0.60,
        ylabel={Average ratio},
        xlabel={Size bin},
        xtick=data,
        symbolic x coords={3--6,7--10,11--40},
        legend style={at={(0.5,1.12)},anchor=south,legend columns=-1},
        ymajorgrids=true,
        grid style={dashed},
        error bars/y dir=both,
        error bars/y explicit,
    ]
    
    \addplot+[draw=black, fill=blue!50] coordinates {
        (3--6, 0.1289) +- (0, 0.0067)
        (7--10, 0.2701) +- (0, 0.0149)
        (11--40, 0.3966) +- (0, 0.0363)
    };
    \addlegendentry{EDVW\_clique}
    
    \addplot+[draw=black, fill=orange!80] coordinates {
        (3--6, 0.1260) +- (0, 0.0065)
        (7--10, 0.2665) +- (0, 0.0125)
        (11--40, 0.3949) +- (0, 0.0392)
    };
    \addlegendentry{EDVW\_hyper}
    
    \addplot+[draw=black, fill=green!60] coordinates {
        (3--6, 0.1337) +- (0, 0.0079)
        (7--10, 0.3052) +- (0, 0.0171)
        (11--40, 0.4560) +- (0, 0.0277)
    };
    \addlegendentry{hyperwalk}
    
    \addplot+[draw=black, fill=red!40] coordinates {
        (3--6, 0.0015) +- (0, 0.0005)
        (7--10, 0.0034) +- (0, 0.0010)
        (11--40, 0.0096) +- (0, 0.0029)
    };
    \addlegendentry{random}
    
    \end{axis}
    \end{tikzpicture}
    \caption{Email-EU dataset performances by size bin for seen hyperedges (mean $\pm$ std across folds), $\alpha=0.5$.}
    \label{fig:TASK2-email-eu-seen-ratio}
\end{figure}


\begin{figure}[h!]
    \centering
    \begin{tikzpicture}
    \begin{axis}[
        ybar,
        width=\linewidth,
        height=0.55\linewidth,
        bar width=8pt,
        enlarge x limits=0.18,
        ymin=0,
        ymax=0.60,
        ylabel={Average ratio},
        xlabel={Size bin},
        xtick=data,
        symbolic x coords={3--6,7--10,11--40},
        legend style={at={(0.5,1.12)},anchor=south,legend columns=-1},
        ymajorgrids=true,
        grid style={dashed},
        error bars/y dir=both,
        error bars/y explicit,
    ]
    
    \addplot+[draw=black, fill=blue!50] coordinates {
        (3--6, 0.2909) +- (0, 0.0165)
        (7--10, 0.2907) +- (0, 0.0083)
        (11--40, 0.3687) +- (0, 0.0314)
    };
    \addlegendentry{EDVW\_clique}

    \addplot+[draw=black, fill=orange!80] coordinates {
        (3--6, 0.2845) +- (0, 0.0152)
        (7--10, 0.2864) +- (0, 0.0087)
        (11--40, 0.3686) +- (0, 0.0304)
    };
    \addlegendentry{EDVW\_hyper}

    \addplot+[draw=black, fill=green!60] coordinates {
        (3--6, 0.3035) +- (0, 0.0174)
        (7--10, 0.3359) +- (0, 0.0132)
        (11--40, 0.4496) +- (0, 0.0290)
    };
    \addlegendentry{hyperwalk}

    \addplot+[draw=black, fill=red!40] coordinates {
        (3--6, 0.0078) +- (0, 0.0039)
        (7--10, 0.0075) +- (0, 0.0016)
        (11--40, 0.0120) +- (0, 0.0015)
    };
    \addlegendentry{random}
    
    \end{axis}
    \end{tikzpicture}
    \caption{Email-EU dataset performances by size bin for novel hyperedges (mean $\pm$ std across folds), $\alpha=0.5$.}
    \label{fig:TASK2-email-eu-novel-ratio}
\end{figure}

\begin{table}[htbp]
    \centering
    \caption{Average theoretical maximum nodes to guess per size bin (across folds) for Email-Eu dataset.}
    \label{tab:avg_theoretical_max_sizebin_email_eu}
    \begin{tabular}{lccc}
    \toprule
    Size bin & Overall & Seen & Novel \\
    \midrule
    2--2        & 8,113 & 8,105 & 8 \\
    3--6        & 14,210 & 12,919 & 1,291 \\
    7--10       & 5,476 & 3,199 & 2,277 \\
    11--$\infty$ & 7,480 & 2,632 & 4,848 \\
    \bottomrule
    \end{tabular}
\end{table}

\subsubsection{Senate-Bills}
\begin{figure}[h!]
    \centering
    \begin{tikzpicture}
    \begin{axis}[
        ybar,
        width=\linewidth,
        height=0.55\linewidth,
        bar width=8pt,
        enlarge x limits=0.18,
        ymin=0,
        ymax=0.35,
        ylabel={Average ratio},
        xlabel={Size bin},
        xtick=data,
        symbolic x coords={2--2,3--6,7--10,11--99},
        legend style={at={(0.5,1.12)},anchor=south,legend columns=-1},
        ymajorgrids=true,
        grid style={dashed},
        error bars/y dir=both,
        error bars/y explicit,
    ]
    
    \addplot+[draw=black, fill=blue!50] coordinates {
        (3--6, 0.0200) +- (0, 0.0105)
        (7--10, 0.0410) +- (0, 0.0037)
        (11--99, 0.1143) +- (0, 0.0189)
    };
    \addlegendentry{EDVW\_clique}
    
    \addplot+[draw=black, fill=orange!80] coordinates {
        (3--6, 0.0328) +- (0, 0.0075)
        (7--10, 0.0498) +- (0, 0.0062)
        (11--99, 0.1909) +- (0, 0.0195)
    };
    \addlegendentry{EDVW\_hyper}
    
    \addplot+[draw=black, fill=green!60] coordinates {
        (3--6, 0.0312) +- (0, 0.0117)
        (7--10, 0.0664) +- (0, 0.0069)
        (11--99, 0.2635) +- (0, 0.0073)
    };
    \addlegendentry{hyperwalk}
    
    \addplot+[draw=black, fill=red!40] coordinates {
        (3--6, 0.0095) +- (0, 0.0036)
        (7--10, 0.0167) +- (0, 0.0019)
        (11--99, 0.0681) +- (0, 0.0018)
    };
    \addlegendentry{random}
    
    \end{axis}
    \end{tikzpicture}
    \caption{Senate-Bills dataset performances by size bin for overall hyperedges (mean $\pm$ std across folds), $\alpha=0.5$.}
    \label{fig:TASK2-senate-bills-overall-ratio}
  \end{figure}
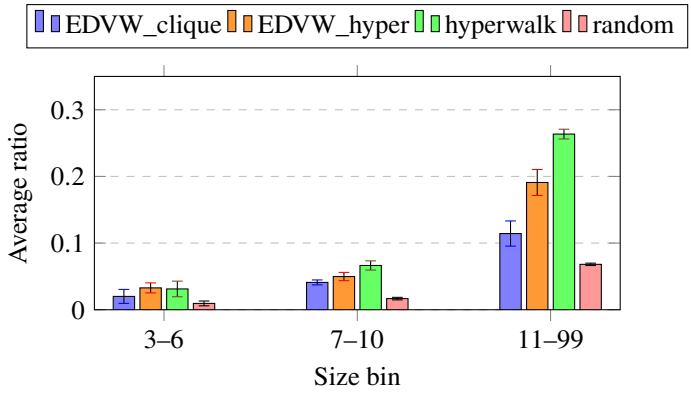

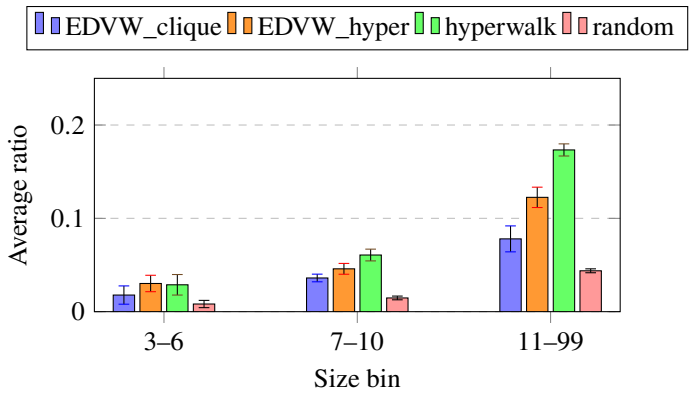
\begin{figure}[h!]
    \centering
    \begin{tikzpicture}
    \begin{axis}[
        ybar,
        width=\linewidth,
        height=0.55\linewidth,
        bar width=8pt,
        enlarge x limits=0.18,
        ymin=0,
        ymax=0.25,
        ylabel={Average ratio},
        xlabel={Size bin},
        xtick=data,
        symbolic x coords={2--2,3--6,7--10,11--99},
        legend style={at={(0.5,1.12)},anchor=south,legend columns=-1},
        ymajorgrids=true,
        grid style={dashed},
        error bars/y dir=both,
        error bars/y explicit,
    ]
    
    \addplot+[draw=black, fill=blue!50] coordinates {
        (3--6, 0.0178) +- (0, 0.0098)
        (7--10, 0.0361) +- (0, 0.0041)
        (11--99, 0.0780) +- (0, 0.0139)
    };
    \addlegendentry{EDVW\_clique}
    
    \addplot+[draw=black, fill=orange!80] coordinates {
        (3--6, 0.0302) +- (0, 0.0088)
        (7--10, 0.0459) +- (0, 0.0058)
        (11--99, 0.1225) +- (0, 0.0109)
    };
    \addlegendentry{EDVW\_hyper}
    
    \addplot+[draw=black, fill=green!60] coordinates {
        (3--6, 0.0288) +- (0, 0.0110)
        (7--10, 0.0607) +- (0, 0.0063)
        (11--99, 0.1733) +- (0, 0.0065)
    };
    \addlegendentry{hyperwalk}
    
    \addplot+[draw=black, fill=red!40] coordinates {
        (3--6, 0.0082) +- (0, 0.0039)
        (7--10, 0.0147) +- (0, 0.0020)
        (11--99, 0.0438) +- (0, 0.0021)
    };
    \addlegendentry{random}
    
    \end{axis}
    \end{tikzpicture}
    \caption{Senate-Bills dataset performances by size bin for seen hyperedges (mean $\pm$ std across folds), $\alpha=0.5$.}
    \label{fig:TASK2-senate-bills-seen-ratio}
  \end{figure}

Here are presented the performances on the \textit{Senate-Bills} dataset of the different combos \textit{Network Architecture + Random Walk} on the \textit{TASK 2} of hyperedge prediction.
In figure ~\ref{fig:TASK2-senate-bills-overall-ratio} the overall ratio of corrects guessed nodes over the theoretical amount 
and ~\ref{fig:TASK2-senate-bills-seen-ratio} of the seen interactions (i.e. it exists at least one identical set \textit{preserved + correctly guessed} that is a subset of a hyperedge in $E^{T}$).
~Table \ref{tab:avg_theoretical_max_sizebin_senate_bills} indicates the average theoretical maximum nodes to guess per size bin (across folds).


\begin{table}[htbp]
    \centering
    \caption{Average theoretical maximum nodes to guess per size bin (across folds) for Senate-Bills dataset.}
    \label{tab:avg_theoretical_max_sizebin_senate_bills}
    \begin{tabular}{lrrr}
    \toprule
    Size bin & Overall & Seen & Novel \\
    \midrule
    3--6        &   667  &   648  &    18 \\
    7--10       & 4,345  & 3,944  &   401 \\
    11--$\infty$ & 96,618 & 35,377 & 61,241 \\
    \bottomrule
    \end{tabular}
\end{table}




\end{document}